\documentclass[twocolumn, twocolappendix, floatfix]{aastex63}
\usepackage{amsmath}

\received{\today}
\revised{\today}
\accepted{???}
\submitjournal{ApJ}

\shorttitle{Small scale dynamo in the ICM}
\shortauthors{Steinwandel et al.}
\graphicspath{{./}{figures/}}

\begin{document}

\title{On the small scale turbulent dynamo in the intracluster medium: A comparison to dynamo theory\footnote{Released on August, 19th, 2021}}

\correspondingauthor{Ulrich P. Steinwandel}
\email{usteinwandel@flatironinstitute.org}

\author[0000-0001-8867-5026]{Ulrich P. Steinwandel}
\affiliation{Center for Computational Astrophysics, Flatiron Institute, 162 5th Avenue, New York, NY 10010}

\author{Ludwig M. B\"oss}
\affiliation{University Observatory Munich, Scheinerstr. 1, D-81679 Munich, Germany}

\author{Klaus Dolag}
\affiliation{University Observatory Munich, Scheinerstr. 1, D-81679 Munich, Germany}
\affiliation{Max Planck Institute for Astrophysics, Karl-Schwarzschildstr. 1, D-85748, Garching, Germany}

\author{Harald Lesch}
\affiliation{University Observatory Munich, Scheinerstr. 1, D-81679 Munich, Germany}



\begin{abstract}
We present non-radiative, cosmological zoom-simulations of galaxy cluster formation with magnetic fields and (anisotropic) thermal conduction of one very massive galaxy cluster with a mass at redshift zero that corresponds to $M_\mathrm{vir} \sim 2 \times 10^{15} M_{\odot}$. We run the cluster on three resolution levels (1X, 10X, 25X), starting with an effective mass resolution of $2 \times 10^8M_{\odot}$, subsequently increasing the particle number to reach $4 \times 10^6M_{\odot}$. The maximum spatial resolution obtained in the simulations is limited by the gravitational softening reaching $\epsilon=1.0$ kpc at the highest resolution level, allowing to resolve the hierarchical assembly of the structures in very fine detail. All simulations presented, have been carried out with the SPMHD-code \textsc{gadget-3} with a heavily updated SPMHD prescription. The primary focus is to investigate magnetic field amplification in the Intracluster Medium (ICM). We show that the main amplification mechanism is the small scale-turbulent-dynamo in the limit of reconnection diffusion. In our two highest resolution models we start to resolve the magnetic field amplification driven by this process and we explicitly quantify this with the magnetic power-spectra and the magnetic tension that limits the bending of the magnetic field lines consistent with dynamo theory. Furthermore, we investigate the $\nabla \cdot \mathbf{B}=0$ constraint within our simulations and show that we achieve comparable results to state-of-the-art AMR or moving-mesh techniques, used in codes such as \textsc{enzo} and \textsc{arepo}. Our results show for the first time in a fully cosmological simulation of a galaxy cluster that dynamo action can be resolved in the framework of a modern Lagrangian magnetohydrodynamic (MHD) method, a study that is currently missing in the literature.
\end{abstract}



\section{Introduction} \label{sec:intro}
Magnetic fields are observed across all scales within the Universe, from the Interstellar medium (ISM) on the scales of molecular clouds \citep[e.g.][]{Clark2014, Clark2019, Crutcher2012, Heiles2005, Sullivan2021, Yue2019} over the large scale field structure in galaxies \citep[e.g.][]{Basu2013, Beck2005, Greaves2000, Jones2020, Lacki2013, Robishaw2008, Tabatabaei2008, Watson2001} to the intra cluster medium (ICM) \citep[e.g.][]{Bonafede2009, Bonafede2010, Bonafede2013, Boehringer2016, Clarke2001, Hu2020}.\\
While there are significant differences in the structure of the magnetic field within the ISM, the IGM and the ICM there seems to be some observational evidence that the magnetic field strength within these very different components is in the order of a few $\mu$G \citep[see e.g.][for reviews on magnetic fields on galaxy, ISM and ICM scales]{Beck2015, Crutcher2012, vanWeeren2019} and seems to be correlated on galaxy and even galaxy cluster scales. \\
Specifically, for galaxy clusters one can determine the magnetic field strength from the Faraday Rotation Measurement (RM) of radio galaxies that are located in the foreground and background of the cluster of interest \citep[e.g.][]{Brentjens2005, Burn1966, Clarke2004, Govoni2004, vanWeeren2019}. The RM measurements of the Coma-galaxy cluster are of the highest quality as they can be obtained from measurements of the magnetic field from seven individual radio galaxies located in the central part of the cluster \citep[see][for the details]{Bonafede2010}. Additionally, the magnetic field of the Coma cluster could be further constrained over seven more radio galaxies detected around the location of the in-falling galaxy group NGC $4839$ \citep{Bonafede2013}. From these studies one can derive a value in the order of a few $\mu$G for the magnetic field in the Coma-cluster. However, the picture may change for cool core clusters where higher magnetic fields of the order of a few 10 $\mu$G are observed \citep[see][for a more detailed review on the observed magnetic field strengths in different galaxy clusters]{vanWeeren2019}. specifically, \citet{Vogt2003} points to cluster magnetic field of around $12$ $\mu$G in the cluster Hydra as a conservative estimate. \\
One can place a lower limit on the magnetic field strength in galaxy clusters which is given by the absence of Inverse Compton emission that should originate from photons that are scattered on cosmic ray electrons in the cluster environment. This should infer a spectrum of hard X-ray emission which could clearly be distinguished from the background thermal Bremsstrahlung spectrum \citep[e.g.][]{Rephaeli1979, Rephaeli1994, Sarazin2000}. However, modelling the Inverse Compton emission is not straightforward \citep[see][and references therein]{vanWeeren2019}. Despite, the difficulty in the modelling the hard X-ray spectrum of Inverse Compton radiation, one can derive an upper limit from which one can infer a lower limit on the magnetic field strength in galaxy clusters ranging from $0.1$ to $0.5$ $\mu$G in the Coma-Cluster \citep[e.g.][]{Rossetti2004}, the Bullet-Cluster \citep[e.g.][]{Wik2014} and Abell $2163$ \citep[e.g.][]{Sugawara2009, Ota2014}.\\
While there seems to be some observational consensus on the magnetic field strengths in galaxy clusters, the situation for the origin of these magnetic fields is less clear. The underlying problem is that the magnetic field amplification in the ICM is supposedly driven by the turbulence that is injected by shocks during the structure formation process \citep[e.g.][]{Miniati2001, Iapichino2012, Iapichino2013, Iapichino2017}. The involvement of ICM MHD-turbulence makes this problem particularly difficult to control, as turbulence is not well understood in numerical simulations to begin with and thus even less as the driver of magnetic field amplification \citep[e.g.][for a very detailed discussion of this problem]{Donnert2018}.\\
Despite the somewhat tedious understanding of turbulence in numerical simulations one can draw a two sided picture of the magnetisation of the ICM. As already pointed out in the classical picture of turbulent amplification of magnetic fields in the ICM, the magnetic field could be amplified during the structure formation process when the largest structures (galaxy clusters) assemble in the Universe. In this framework the idea is that the magnetic field is amplified due to turbulence in the ICM driven by cosmic accretion that drives strong shocks in the ICM. This is a very intriguing picture because despite the reality of high Mach number shocks in the ICM \citep[e.g.][]{Miniati2001}, turbulence in the ICM is vastly sub-sonic, given the high temperature of around 10$^{8}$ K in the ambient medium. Thus magnetic field amplification in the ICM could be described beautifully by the dynamo theory first developed by Kazantsev and Kraichnan \citep{Kazantsev1968, Kraichnan1967} which has been further advanced by several authors since then \citep[e.g.][]{Boldyrev2004, Kazantsev1985, Kulsrud1992, Kulsrud1997, Ruzmaikin1988, Subramanian2002, Xu2020, Zeldovich1965, Zeldovich1970, Zeldovich1983}. \\
This theory self-consistently describes the amplification of magnetic fields via sub-sonic turbulence due to stretching, twisting and subsequent folding of magnetic field lines and can be tested through means of the magnetic power spectrum and the distribution of the curvature of magnetic field lines derived from high resolution numerical simulations \citep[e.g.][]{Schekochihin2004, Porter2015}. This theory has been widely applied on the scales of galaxies \citep[e.g.][]{Wang2009, Kotarba2009, Beck2012, Pakmor2013, Martin2020, Marinacci2015, Marinacci2016, Pakmor2017, Pakmor2020, Rieder2016, Rieder2017a, Rieder2017b, Steinwandel2019, Steinwandel2020, Steinwandel2020b}. While the concept of turbulent magnetic field amplification works well on galaxy scales, the ideal application for this theoretical framework is magnetic field amplification in the ICM due to the sub-sonic nature of ICM-turbulence. \\
Already in early numerical simulations of magnetic fields in galaxy clusters it has been pointed out that the amplification via the small-scale-turbulent-dynamo is quite likely without showing direct evidence of the process \citep{Brueggen2005, Dolag1999, Dolag2001, Dolag2002, Dolag2005, Dubois2008, Ryu2008, Vazza2014} and recently the first efforts have been made to show direct evidence of an acting small-scale-turbulent dynamo on the scales of galaxy-clusters in Eulerian codes \citep[e.g.][]{Vazza2018, Rho2019} by directly comparing to the statistics that is enforced by dynamo theory and high resolution numerical simulations of the small-scale turbulent dynamo in the ICM-regime \citep[e.g.][]{Schekochihin2004, Porter2015}. \\
Generally, the idea of the small-scale turbulent dynamo is that the magnetic field is generated on the scales of the turbulent eddies and is thus of scale-free nature. That means the stretching, twisting and folding of field lines can occur on parsec (pc) scales in the ISM or on megaparsec (Mpc) scales in the ICM with a growth rate that is proportional to the eddy-turn-over time with respect to the ambient medium. The field is then propagated to the larger-scales via an inverse turbulence cascade when the magnetic energy density reaches equipartition with the turbulent energy density stored in the smallest eddies. This process leads to an increase of the power stored in the magnetic field on the larger scales with a subsequent decrease on the largest scales that is predicted by the evolution of the energy spectra in Kazantsev-Kraichnan theory.\\
In this paper we study the build-up of the magnetic field in numerical simulations based on the theory of the small-scale-turbulent dynamo in the framework of a Lagrangian simulation framework that is currently entirely missing within the literature of magnetic field amplification in galaxy clusters.\\
The paper is structured as follows. In section \ref{sec:theory} we discuss the theoretical background that is needed to understand the basics of the theory of the small scale turbulent dynamo. In section \ref{sec:numerics} we discuss the basics of the numerical algorithms we use to carry out our simulations, including the handling of non-ideal MHD. In section \ref{sec:ICS} we discuss the details of the simulation suite alongside the initial conditions and the adopted naming conventions. In section \ref{sec:results} we present the results of the simulations and carry out the analysis that is needed to study the turbulent dynamo in the ICM. In section \ref{sec:conclusions} we summarise our findings, conclude our results and comment on model limitations and future work. Furthermore, we will discuss model variations in the Appendix of this work.
 
\section{Short overview on small-scale-turbulent dynamo theory \label{sec:theory}}
Magnetic field amplification in the ICM is most likely driven by turbulence that is injected during structure formation. The idea is that tiny magnetic fields of which the origin is still under debate \citep[e.g.][]{Biermann1950, Demozzi2010, Matarese2004, Gnedin2000, Rees1987, Rees1994, Rees2005, Rees2006} are amplified to large-scale coherent fields via stretching, twisting and folding of magnetic field lines. This process is limited by the magnetic tension force that makes every stretch-twist-fold process inharently more difficult and finally saturates the dynamo when the turbulent kinetic energy in the smallest eddies is in equipartition with the magnetic field energy. Over the past sixty years several authors have continuously refined the theoretical understanding of turbulent magnetic field amplification \citep[e.g.][to just name a few]{Boldyrev2004, Kraichnan1967, Kazantsev1968, Kazantsev1985, Kulsrud1992, Kulsrud1997, Ruzmaikin1988, Subramanian2002, Zeldovich1965, Zeldovich1970, Zeldovich1983, Xu2020}. The picture of turbulent dynamo amplification fits perfectly to the use case of magnetic field amplification in galaxy clusters as the ICM is very hot and therefore is characterised by the subsonic turbulent energy cascade that is in very good agreement with Kolmogorov theory of turbulence \citep{Kolmogorov1941}. \\
Every theory of magnetic field amplification starts with the induction equation in the continuum limit of MHD: 
\begin{align}
    \frac{\partial \mathbf{B}}{\partial t} = \nabla \times (\nabla \times \mathbf{B}) + \eta \Delta \mathbf{B},
\end{align}
where $\eta$ is the magnetic diffusivity. 
A quite intuitive approach towards an understanding of the magnetic field structure developed by the small-scale turbulent dynamo can be derived by statistically studying fluctuations in velocity and magnetic field. A general vector field can always be Fourier decomposed. Thus, this means the velocity field can be written as $\mathbf{v} = \int d^3k \mathbf{k} \mathbf{v}_{\mathbf{k}} \cdot e^{i\mathbf{k} \cdot \mathbf{r}}$ and the magnetic field can be written as $\mathbf{B} = \int d^3k \mathbf{k} \mathbf{B}_{\mathbf{k}} \cdot e^{i\mathbf{k} \cdot \mathbf{r}}$. The ultimate goal of the Fourier analysis of these fluctuations is to derive the distribution of the power in the magnetic field that is given via:
\begin{align}
    E_\mathrm{mag} = \frac{<\mathbf{B}^2>}{8 \pi} = \int P_\mathrm{M}(k) dk.
    \label{eq:mag_dens}
\end{align}
One can find the time-derivative of $P_\mathrm{M}(k)$ \citep[see.][for details of the derivation]{Kulsrud2008}:
\begin{align}
    \frac{\partial P_\mathrm{M}(k)}{\partial t} = \int K(k,k_{0}) M(k_{0}) dk_{0} - 2\beta k^2 P_\mathrm{M}(k),
    \label{eq:mag_struct}
\end{align}
which describes the time evolution of the magnetic power spectrum as a function of structure function $K$ and turbulent resistivity $\beta$. The combination of equation \ref{eq:mag_dens} with \ref{eq:mag_struct} yields:
\begin{align}
  \frac{dE_\mathrm{mag}}{dt} = 2 \gamma E_\mathrm{mag},  
\end{align}
where $\gamma$ denotes the growth-rate of the dynamo. From this one can straight forward see that the magnetic field strength is increased by a factor of two as a function of the eddy-turn-over time. This is consistent with the stretching, twisting and folding as it is assumed to occur on the small-scale turbulent dynamo. Furthermore, it is worth noting that in this prescription the growth rate is then directly correlated with the eddy-turn-over-rate of the smallest eddies and energy is carried to the larger scales by an inverse turbulent cascade. In the kinematic regime one can find $P_\mathrm{M}(k)$ by solving:
\begin{align*}
    \frac{\partial P_\mathrm{M}(k)}{\partial k} &= \frac{\gamma}{5} \left(k^2\frac{\partial^2 P_\mathrm{M}(k)}{\partial k^2} - 2 k\frac{\partial P_\mathrm{M}(k)}{\partial k} + 6 P_\mathrm{M}(k)\right)
\end{align*}
\begin{align}
    &- 2k^2 \lambda_\mathrm{res} P_\mathrm{M}(k), 
\end{align}
where $\lambda_\mathrm{res}$ is the resistivity. This differential equation can be solved with standard methods and one obtains:
\begin{align}
    P_{M}(k,t) \propto e^{3/4 \gamma t} k^{3/2},
\end{align}
which directly indicates exponential growth of modes with $k^{3/2}$. Therefore, the small-scale-turbulent dynamo can be clearly identified over the shape of the magnetic energy spectra. This brief estimate shows why many groups investigate the power spectrum to identify the dynamo. However, the shape of the power spectra in numerical simulations is often very generic and it remains unclear what is driving the shape of the power spectrum. We think this is an important point and therefore one consider the following example to understand why identifying the dynamo by the power spectrum alone might be problematic. In supersonic turbulence one often considers Burgers turbulence with a power law of the form $k^{-2}$. One can quite easily derive the power spectrum that is inferred by series of shock waves. This will yield the same power law of the form $k^{-2}$. Therefore, this raises the question when is turbulence and when are shocks the origin of this behaviour. The answer is that one can use the density PDF to distinguish the two as shock waves will naturally lead to deviations from the log-normal PDF known from supersonic turbulence\footnote{We are aware that this is strictly speaking only true for a non-gravitating fluid}.\\ 
Thus, why would the one identify the dynamo only over the shape of a power-spectrum, which is also tedious to obtain (in Lagrangian methods at least)? From theoretical calculations it is inferred that the stretching twisting and folding of magnetic field lines is limited by the magnetic tension force which inevitably generates an imprint on the bending of the field lines itself. Therefore, another popular way to test dynamos in numerical simulations is to test the dependence of the magnetic field strength on the curvature of a related field line \citep[e.g.][]{Schekochihin2004, Vazza2018, Steinwandel2019}.

\section{Numerical Method} \label{sec:numerics}
We carry out the simulations presented in this paper with the Tree-SPMHD-Code \textsc{p-gadget3} which is the developers version of the Tree-SPH-Code \textsc{p-gadget2} \citep{Springel2005}. We use a modern implementation of SPH, as presented in \citet{Beck2016} that includes time-dependent artificial viscosity and conduction and employs higher order kernel functions, given as the well studied Wendland functions \citep{Wendland1995, Wendland2005, Dehnen2012}. However, as thermal conduction is very important in the ICM we do not use the-time dependent artificial conduction implementation within the simulations but rather use the physical conduction implementation first presented in \citet{Jubelgas2004} and later updated by \citet{Arth2014} to a conjugate gradient solver for improved convergence and stability of the scheme. The conjugate gradient solver employed is similar to the one developed in \citet{Petkova2009} for the use in galaxy formation simulations that include direct radiative transfer. We run all the simulations with physical conduction and $1/20$th of the canonical Spitzer-value. Our version of \textsc{p-gadget3} further includes magnetic fields and magnetic dissipation as presented in \citet{Dolag2009}. While we will use isotropic thermal conduction in our default runs in the main paper we carry out some additional runs utilising anisotropic thermal conduction following the prescription in \citet{Arth2014}. We note that the usage of anisotropic thermal conduction is increasing the computational cost of a simulation by roughly $20$ per cent. In the following, we briefly discuss the specifics of the underlying SPH-equations.

\subsection{Kernel function and density estimate}

The current SPH-scheme implemented in our code is based on the density-entropy formulation of SPH which means that we smooth the density field in the following fashion:
\begin{align}
  \rho_{i} = \sum_{j} m_{j} W_{ij}(x_{ij}, h_{i}), 
  \label{eq:sph}
\end{align}
where $h_{i}$ is the smoothing-length. The summation carried out in equation \ref{eq:sph} is computed over the neighbouring particles within the kernel $W_{ij}(x_{ij}, h_{i})$:
\begin{align}
  W_{ij}(x_{ij}, h_{i}) = \frac{1}{h_{i}^3} w(q). 
\end{align} 
In our simulations the kernel function is used with 295 neighbouring particles. The function $w(q)$ is given by  
\begin{align}
  w(q) = \frac{1365}{64 \pi} (1-q)^6 \left(1 + 8q + 25 q^2 + 32 q^3\right), 
\end{align}
for $q<1$. For $q>1$ we set $w(q)$ to zero and is known as the Wendland C6 kernel function \citep[][]{Wendland1995, Wendland2005, Dehnen2012}.

\subsection{Equation of motion in SPH and SPMHD}

The equations of motion (EOM) for SPH can conveniently be derived from a discrete Lagrangian as presented in \citet{Price2012} by using the physical principle of least action. This has the distinct advantage that the derived formulation is conserving energy, momentum and angular momentum by construction. This leads to the SPH-formulation of the EOM in the pure hydrodynamic case:
\begin{align}
  \frac{\mathrm{d} \mathbf{v}_{i}}{\mathrm{d}t} = -\sum_{j} m_{j} \left[f_{i}^\mathrm{co} \frac{P_{j}}{\rho_{j}^{2}} \frac{\partial W_{ij}(h_{i})}{\partial \mathbf{r}_{i}} + f_{j}^\mathrm{co} \frac{P_{j}}{\rho_{j}^{2}} \frac{\partial W_{ij}(h_{j})}{\partial \mathbf{r}_{i}} \right].
\label{eq:EOMsph}
\end{align}   
with $f_{j}^\mathrm{co}$ given by
\begin{align}
  f_{j}^\mathrm{co} = \left[1 + \frac{h_{j}}{3 \rho_{j}} \frac{\partial \rho_{j}}{\partial h_{j}}\right]^{-1}.
\end{align}
A similar argument can be made for the SPMHD case leading to the SPH formulation of the MHD EOM given as:
\begin{align}
  \frac{\mathrm{d}\mathbf{v}_{i}}{\mathrm{d}t} = -\sum_{j} m_{j} \left[f_{i}^\mathrm{co} \frac{P_{i} + \frac{1}{2\mu_{0}} B_{i}^{2}}{\rho_{i}^{2}} \nabla_{i} W_{ij}(h_{i}) \right. \notag \\
  \left. + f_{j}^\mathrm{co} \frac{P_{j} + \frac{1}{2\mu_{0}} B_{j}^{2}}{\rho_{j}^{2}}  \nabla_{i} W_{ij}(h_{j}) \right] \notag \\
      + \frac{1}{\mu_{0}} \sum_{j} m_{j} \left[f_{i}^\mathrm{co} \frac{\textbf{B}_{i}[\textbf{B}_{i} \cdot \nabla_{i} W_{ij}(h_{i})]}{\rho_{i}^{2}} \right. \notag \\
      \left. + f_{j}^\mathrm{co}\frac{\textbf{B}_{j}[\textbf{B}_{j} \cdot \nabla_{i} W_{ij}(h_{j})]}{\rho_{j}^{2}}\right]. 
  \label{eq:SPMHD}
\end{align}
The presence of the magnetic field alters the EOM in several ways. First, the presence of the magnetic field is leading to an additional pressure component, apart from the thermal pressure within the fluid. This additional pressure component scales as $\mathbf{B^2}$. The fact that the magnetic pressure component scales as $\mathbf{B^2}$ is crucial to establish pressure equipartition with the thermal pressure relatively quickly. Second, the term on the left hand side of equation \ref{eq:SPMHD} is arising due to the divergence cleaning constraint $\nabla \cdot \textbf{B}=0$. This term is problematic because it breaks the symmetry of the underlying Lagrangian in a way that the system is not invariant under rotation of the system anymore. Thus, from Noether's theorem one can easily see that the SPMHD-equations are not strictly conserving angular momentum.

\subsection{Formulation of the induction equation in non-ideal MHD with effective $\eta_\mathrm{m}$}

Furthermore, it is not only interesting how the magnetic field is influencing the EOM but also how the magnetic field itself is evolving with time. The evolution of the magnetic field is generally given by the induction equation that takes the form:
\begin{align}
    \frac{\partial \mathbf{B}}{\partial t} = \nabla \times \left( \mathbf{v} \times \mathbf{B}\right) + \nabla \times \eta_\mathrm{m} (\nabla \times \mathbf{B})
\end{align}
which can be re-formulated as:
\begin{align}
     \frac{\partial \mathbf{B}}{\partial t} = (\mathbf{B} \cdot \nabla)\mathbf{v} - \mathbf{B} (\nabla \cdot \nabla) + \eta_\mathrm{m} \Delta \mathbf{B}. 
     \label{eq:non_ideal}
\end{align}
To this day most MHD simulations of galaxies or galaxy-clusters drop the last term of equation \ref{eq:non_ideal} and there are only very few simulations that include these terms \citep[e.g.][]{Kotarba2011, Bonafede2011, Steinwandel2019, Steinwandel2020}. However, this term is crucial for modelling the plasma in an accurate fashion\footnote{Essentially, there are no dynamos that properly work without some form of diffusion} and thus we include it in our cosmological galaxy cluster simulations. The parameter $\eta_\mathrm{m}$ is hereby an effective diffusion parameter that is comprised of the contribution due to thermal conduction by $\eta_\mathrm{coulomb}$ that is related to the thermal conductivity $\sigma$ and the turbulent diffusion coefficient of the plasma $\eta_\mathrm{turb}$. While the exact value of the diffusion coefficient in the ICM is under debate and several processes yield different limits \citep[e.g.][]{Strong2007, Lesch2003, Schlickeiser1987, Schuecker2004, Maier2009, Rebusco2006} we use a moderate value of $\sim \eta_\mathrm{turb} \sim 2 \cdot 10^{27}$ cm$^2$ s$^{-1}$. This is in good agreement with the classical Spitzer-model for the ICM \citep{Spitzer1956}. For a more detailed discussion on the choice of $\eta_\mathrm{m}$ we refer to section 4.2 of \citet{Bonafede2011}. \\
Following \citet{Dolag2009} the diffusion term in the induction equation takes the form:
\begin{align}
    \frac{\partial \mathbf{B}}{\partial t} = \frac{\eta_\mathrm{m}\rho_{i}}{Ha^2} \sum_{j} = \frac{m_{j}}{\rho_{j}} (\textbf{B}_{i} - \mathbf{B}_{j}) \frac{(\mathbf{r}_{i} - \mathbf{r}_{j})}{|(\mathbf{r}_{i} - \mathbf{r}_{j})|} \cdot \nabla_{i} W_{i}, 
\end{align}
where $W$ is the kernel and $\mathbf{B}_{i}$ and $\mathbf{B}_{j}$ are the magnetic field vectors at the positions $\mathbf{r}_{i}$ and $\mathbf{r}_{j}$. $Ha^2$ is the internal scaling applied in \textsc{gadget} needed for correct unit conversion from the co-moving field $\mathbf{B}$ to the physical field $\mathbf{B}/a^2$. However, as the magnetic field is dissipated this introduces an additional entropy term for the thermal plasma, manifesting in the rate of change of entropy:
\begin{align}
    \frac{dA}{dt} = -\frac{\gamma - 1}{2 \mu_{0} \rho_{i}^{\gamma - 1}} \sum_{j} \frac{m_{j}}{\rho_{j}} (\textbf{B}_{i} - \mathbf{B}_{j})^2 \frac{(\mathbf{r}_{i} - \mathbf{r}_{j})}{|(\mathbf{r}_{i} - \mathbf{r}_{j})|} \cdot \nabla_{i} W_{i}.
\end{align}
Further implementation details can be found in \citet{Dolag2009}.

\subsection{Thermal conduction}

Thermal conduction is believed to be a process of importance in the ICM. We will briefly describe the physical process as well as the numerical implementation of the process into our simulation code for the isotropic and the anisotropic case. We note that while we run most of the simulations with isotropic conduction in the presence of magnetic fields we run one simulation with anisotropic conduction and show the results in Appendix \ref{appendix:C}. We note that at our adopted $5$ per cent of the Spitzer value the difference between the isotropic case is only marginal and we adopted isotropic thermal conduction to safe computational cost. 

\subsubsection{The isotropic case}
Thermal conduction is the physical processes that describes heat transfer via scattering of free electrons. Thus in order to properly work one needs a high ionisation fraction of the underlying plasma. In the ICM this is straightforward achieved as the ICM has virial temperatures the can easily reach $10^8$ K. One can follow \citet{Spitzer1956} to get the the heat flux as a function of the gradient of the temperature distribution via:
\begin{align}
    \mathbf{Q} = - \kappa \nabla T, 
\end{align}
where $\kappa$ is the conduction coefficient. In the classical Spitzer case one makes the assumption of an idealised Lorentzian gas for which one can find the canoncial Spizter-value given by:
\begin{align}
    \kappa_{Sp} = 20 \left(\frac{2}{\pi}\right)^{3/2} \frac{\left(k_\mathrm{B} T_\mathrm{e}\right)^{5/2}}{m_\mathrm{e}^{1/2} e^{4} Z \ \ln \Lambda},
\end{align}
where k$_\mathrm{B}$ is the Boltzmann constant, m$_\mathrm{e}$ is the electron mass, T$_\mathrm{e}$ is the electron temperature $e$ is the elementary charge, Z is the the average number of protons in the plasma and $\ln$ $\Lambda$ is the Coulomb logarithm. As the temperature of clusters is very high we and the conduction coefficient shows a strong dependence with the temperature one can infer a strong contribution of heat conduction to the dynamical processes driven in the ICM. In a realistic plasma one typically does not reach the Spitzer-regime and conduction is suppressed. This is strongly dependent on the average number of protons present in the plasma \citep[see][]{Spitzer1953} which results in a suppression factor of $0.3$ \citep[][]{Spitzer1956} under the assumption of a primordial distribution of the gas, which is a good first order assumption for cosmological simulations. However, one often applies the following parameterization of $\kappa$ in cosmological simulations:
\begin{align}
    \kappa = 1.31 \cdot n_\mathrm{e} \lambda_\mathrm{e} k_\mathrm{B} \left(\frac{k_\mathrm{B} T_\mathrm{e}}{m_\mathrm{e}}\right)^{1/2},
\end{align}
which can be refactored to obtain:
\begin{align}
    \kappa = 4.6 \cdot 10^{13} \left(\frac{T_\mathrm{e}}{10^8 \mathrm{K}}\right)^{5/2} \frac{40}{\ln \Lambda} \frac{\mathrm{erg}}{\mathrm{s \ cm \ K}},
\end{align}
where n$_{e}$ is the electron number density and $\lambda_\mathrm{e}$ is their mean free path. From this one can directly see why the heat transport is dominated by electrons or why the heat transport by protons in the plasma is sub-dominant, as the conductivity scales with the inverse mass. Thus heat conduction is dominated by the electron population of the plasma\footnote{The inverse mass is sometimes referred to as the mobility when re-scaled with the mean free path of the particle.}. Thus we neglect any contribution of protons and furthermore make the assumption that the Coulomb logarithm is constant with $\ln \Lambda =37.8$. This picture is incomplete as for now we have assumed that the temperature gradient which is given via:
\begin{align}
    l_\mathrm{T} =\frac{T}{|\nabla T|},
    \label{eq:cond}
\end{align}
is much larger than the mean free path of the electrons. Strictly speaking one can only make this assumption in a higher density plasma. However, in a lower density plasma such as the ICM where the temperature gradient and the mean free path of the electrons are roughly of the same order transporting energy by conduction is limited by low number of interaction rates in the plasma. Thus we are in the conduction-saturation limit of \citet{Cowie1977} who computed the limited heat flux for a low density plasma:
\begin{align}
    Q_\mathrm{sat} = 0.4 n_\mathrm{e} k_\mathrm{B} T \left(\frac{2 k_{B} T}{\pi m_\mathrm{e}}\right)^{1/2}.
    \label{eq:cond_sat}
\end{align}
Now, one can interpolate between equation \ref{eq:cond} and equation \ref{eq:cond_sat} to obtain the total heat flux:
\begin{align}
    Q_\mathrm{tot} = -\frac{\kappa T}{l_\mathrm{T} + 4.2 \lambda} \frac{\nabla T}{|\nabla T|}.
\end{align}
This is equivalent to a re-normalised conduction coefficient:
\begin{align}
    \kappa = \frac{\kappa{Sp}}{1+4.2\lambda/l_\mathrm{T}}. 
\end{align}
Therefore, one can finally formulate the rate of change of the energy per unit mass:
\begin{align}
    \frac{du}{dt} = -\frac{1}{\rho} \nabla \cdot \mathbf{Q} = \frac{1}{\rho} \nabla \cdot (\kappa \nabla T).  
\end{align}

\subsubsection{The anisotropic case}

In a magnetised plasma heat conduction is slightly more complicated because as the electrons are charged particles their scattering processes are influenced by the magnetic field structure. While electrons can move freely alongside magnetic field lines their motion perpendicular to the field lines is suppressed. Hence the term anisotropic conduction. The trajectory of electrons in the presence of magnetic fields is well studied and the gyrate around magnetic field lines with the Larmor-frequency:
\begin{align}
    \omega_\mathrm{g} = \frac{eB}{m_\mathrm{e} c},
\end{align}
where $c$ is the speed of light. This affects the movement of the electrons in the presence of magnetic fields as pointed out by \citet{Frank-Kamenezki1967}. Following \citet{Braginskii1965} one can subdivide the heat flux in three additive terms
\begin{align}
    \mathbf{Q} = - \kappa_{||} \nabla_{||} T - \kappa_{\perp} \nabla_{\perp} T - \kappa_{\Lambda} \mathbf{B} \times \nabla T, 
\end{align}
where the first two are referred to as the parallel and the perpendicular component and the last term is called the 'Hall-term'. We drop this term as we will see that is will vanish once we start the discretization of our numerical scheme. However, the remaining contributions are extremely tedious to describe. To do so it is often useful to introduce so-called diffusion coefficients $D$ that are related to the conduction coefficient $\kappa$ by $\kappa \sim D n_\mathrm{e} k_{\mathrm{B}}$. One cane now distinguish between two cases for the perpendicular diffusion coefficient. First, a diffusion coefficient that scales as B$^{-2}$. Second, a diffusion coefficient that scales with B$^{-2}$. \\
In the former case one can assume that the diffusion coefficient $D$ can be described as $D \approx v^2 \tau$. Since electrons can move freely along the field lines we get $D_{||}=D$. On the other hand perpendicular to the field liens electrons can only move by interchanging cyclotron frequencies so that $D_{\perp} \approx \lambda_{v}/(\omega_\mathrm{g}^2 \tau^2)$ and one can straightforward determine that $D_{\perp}/D_{||} \propto B^{-2}$ for $\omega_\mathrm{g} \tau \ll 1$. If the gyro-radius is of the order of the mean free path one finds the relation $D_{\perp}/D_{||} \approx 1/(1+\omega_\mathrm{g}^2\tau)$. For typical values of the ICM one obtains $D_{\perp}/D_{||} \approx 10^{-28}$.\\
However, in practice it is a bit more complicated because the transport process perpendicular to the field lines is interacting with turbulent diffusion processes and possibly reconnection diffusion events which make the interaction highly non-linear and experiments conducted in the laboratory indicate that the scaling is rather of order B$^{-1}$ than B$^{-2}$ given by Bohm diffusion \citep[][]{Guthrie1949}. \\
Finally, we can write down the heat flux in the case of anisotropic thermal conduction:
\begin{align}
    \mathbf{Q} = -\kappa \frac{\nabla T \cdot \mathbf{B}} {|\nabla T \cdot \mathbf{B}|} \nabla T =: -\kappa \cos(\theta) \nabla T.
\end{align}
In practice we will split the conduction equation in two parts which results in the rate of change of the energy per unit mass of the form
\begin{align}
    \frac{du}{dt} = \frac{1}{\rho} \nabla \cdot \left[\kappa_{||} \left(\mathbf{\hat{B}} \cdot \nabla T\right) \mathbf{\hat{B}} + \kappa_{\perp} \left(\nabla T -\left(\mathbf{\hat{B}} \nabla T\right)\mathbf{\hat{B}}\right)\right].
    \label{eq:not_dis}
\end{align}

\subsubsection{The numerical implementation}

Now we need to discretize equation \ref{eq:not_dis} to obtain the its SPH-formulation. First, we will re-write equation \ref{eq:not_dis} by factoring ($\mathbf{\hat{B}} \cdot \nabla T) \mathbf{\hat{B}}$ which yields:
\begin{align}
    \frac{du}{dt} = \frac{1}{\rho} \nabla \cdot \left[ (\kappa_{||} - \kappa_{\perp}) \left(\mathbf{\hat{B}} \cdot \nabla T\right) \mathbf{\hat{B}} + \kappa_{\perp} \nabla T\right].
    \label{eq:dis}
\end{align}
A straightforward way to solve this equation is to do it operator-split and solve for the divergence and the temperature gradient in chained SPH loops. In practice, this has the disadvantage that it increases the computational cost and leads to increased noise in the solution as each loop will independently add partition noise to the final result. Thus this is not the favoured way of solving this problem. We will follow the methodology derived in \citet[][]{Petkova2009} who discretized a similar diffusion equation in the context of radiative transfer. In the following we will derive the discrete form of the anisotropic conduction equation but only for the first term. The reason for this is that the second term in equation \ref{eq:dis} can be solved as described in \citet{Jubelgas2004} or \citet{SteinwandelSNpaper} and thus we refer to those papers for a review of how to solve the second term in SPH. The first term takes the discrete form of
\begin{align}
    \left(\frac{du}{dt}\right)^{1\mathrm{st}} = \frac{1}{\rho}\sum_{\alpha, \beta} \frac{\partial}{\partial x_{\alpha}}\left[(\kappa_{||} - \kappa_{\perp}) \hat{B}_{\alpha} \hat{B}_{\beta} \frac{\partial}{\partial x_{\beta}}T\right]. 
\end{align}
We note that $alpha$ and $\beta$ represent the components of a tensor of second order. We substitute the tensor components by $A_{\alpha \beta} = (\kappa_{||} - \kappa_{\perp}) \hat{B}_{\alpha} \hat{B}_{\beta}$, which yields
\begin{align}
    \left(\frac{du}{dt}\right)^{1\mathrm{st}} = \frac{1}{2 \rho} \sum_{\alpha, \beta}\left[\frac{\partial^2 A_{\alpha \beta} T}{\partial x_{\alpha} \partial x_{\beta}} - T \frac{\partial^2 A_{\alpha \beta} }{\partial x_{\alpha} \partial x_{\beta}} + A_{\alpha \beta} \frac{\partial^2 T}{\partial x_{\alpha} \partial x_{\beta}}\right].
    \label{eq:second_derivative}
\end{align}
Now we want to re-write the second derivative present in equation \ref{eq:second_derivative}. This can be achieved by using the following identity for an arbitrary vector $\mathbf{Y}$ \citep[e.g.][]{Price2012}. 
\begin{align}
    \sum_{\alpha} \frac{\partial^2 Y_{i}}{\partial x_{\alpha}^2} = 2 \int d^3 x_{j} (Y_{j} -Y_{i}) \frac{\mathbf{x_{ij}^{\mathrm{T}}} \cdot \nabla_{i} W{ij}}{|\mathbf{x}_{ij}|^2}.
    \label{eq:identity}
\end{align}
We can use equation \ref{eq:identity} to re-write the right hand side of equation \ref{eq:second_derivative} to obtain
\begin{align}
    \left(\frac{du}{dt}\right)^{1\mathrm{st}} = \frac{1}{\rho_{i}} \int d^3 \mathbf{x}_{i} \mathbf{x}_{j}^{T} \left[\frac{(\mathbf{A}_{j} + \mathbf{A}_{i}) (T_{j} - T_{i})}{|\mathbf{x}_{ij}|^2}\right] \nabla_{i} W_{ij}.
    \label{eq:next}
\end{align}
Finally, we can write the integral on the right hand side of equation \ref{eq:next} by the SPH-sum over the neighbours to get
\begin{displaymath}
    \left(\frac{du}{dt}\right)^{1\mathrm{st}} = \ \ \ \ \ \ \ \ \ \ \ \ \ \ \ \ \ \ \ \ \ \ \ \ \ \ \ \ \ \ \ \ \ \ \ \ \
\end{displaymath}
\begin{align}
    \frac{\mu(\gamma -1)}{k_\mathrm{B} \rho_{i}} \sum_{j}^{N_\mathrm{ngb}} \frac{m_{j}}{\rho_{j}} \cdot \mathbf{x}_{ij}^{\mathrm{T}} \left[\frac{(\mathbf{A}_{j} + \mathbf{A}_{i}) (u_{j} - u_{i})}{|\mathbf{x}_{ij}|^2}\right] \nabla_{i} W_{ij},
    \label{eq:finl_dis}
\end{align}
where $\mu$ is the molecular weight and $\gamma$ is the adiabatic index which we set to $5/3$ in the whole simulation domain. We note that we essentially recover the same scaling of the anisotropic case with the temperature difference between SPH particles as we know it from \citet{Jubelgas2004} and \citet{SteinwandelSNpaper}. The problem with equation \ref{eq:finl_dis} is that there is a pre-condition for actually solving this for the tensor $\mathbf{A}_{i} + \mathbf{A}_{j}$ which needs to be positive definite to establish physical heat flux from hot to cold. This is related to the fact that $A$ scales with the difference between $\kappa_{||}$ and $\kappa_{\perp}$, which can be negative for strong anisotropies of the heat flux. Thus technically the heat flux can be negative and flow from cold to hot which is nonphysically. There are several methods to circumvent this. First, one could essentially take the approach of flux limited diffusion. Second, only do the anisotropic part if the tensor is positive definite. Third, isotropise the tensor. We follow \citet{Petkova2009} and do the latter by adding an artificial isotropic component to the heat flux tensor that yields $\mathbf{A} \rightarrow \alpha \mathbf{A} + 1/3 (1-\alpha) \mathrm{tr}(\mathbf{A})\mathbf{1}$. We follow \citet{Petkova2009} and adopt $\alpha=2/3$. We note that we actually do not do a discretisation of the 'Hall-term' since in our parameterisation it is simply vanishing from the discretised equations. \\ 
Finally, we solve the actual differential equation. We do this by adopting the method from \citet{Petkova2009} with the so-called conjugate gradient method, which requires an additional SPH loop but shows very accurate results in practice. The conjugate gradient method is in principle a numerical method to solve a matrix inversion problem of the form 
\begin{align}
    \mathbf{C} \cdot{x} = \mathbf{b}.
\end{align}
The method is implicit and iterates to a solution with a very high convergence order. However, it is numerically unstable, which means that it does not guarantee a solution for arbitrary particle distributions. However, the underlying idea of a conjugate gradient method is the following. The algorithm is supposed to monotonically approach the solution of each element of the inverted matrix by adopting a weight that is dependent on the residual of the previous iterations based on a good initial guess of the state of the physical system. In order for this to work $\mathbf{C}$ needs to be real, positive definite. However, as we already use a correction to get an isotropic version of our heat flux tensor we can assume that both these assumptions are valid. The final task that remains is to write  equation \ref{eq:finl_dis} in the form of conjugate gradient
\begin{align}
    u_{i}^{n+1} = u_{i}^{n} + \sum_{j}^{N_\mathrm{ngb}}, c_{ij}(u_{i}^{n+1} - u_{j}^{n+1})
\end{align}
where we adopted $du_{i}/dt  = \Delta u/\Delta t = (u_{i}^{n+1} - u_{i}^{n})/\Delta t$. The $c_{ij}$ are computed as follows
\begin{align}
    c_{ij} = -\frac{(\gamma -1) \mu}{k_\mathrm{B}} \frac{m_{j} \Delta t}{\rho_{i} \rho_{j}} \frac{\mathbf{x}_{ij}^{\mathrm{T}}}{|\mathbf{x}_{ij}|^2} \left(\mathbf{A}_{i} + \mathbf{A}_{j}\right) \nabla_{i} W_{ij}.
\end{align}
This is the conjugate gradient version of our anisotropic thermal conduction equation (the first term). We noted that the scheme can be numerically unstable. However, this can be resolved by switching to a bi-conjugate method (\textsc{bicgstab}) which applies convergence through a second direction (geometrically speaking). \\
We note that there are a lot of similar implementations that treat different parts of thermal conduction mostly in the realm of galaxy cluster formation \citep[][]{Ruderman2000, Dolag2004, Schekochihin2008, Rasera2008, Parrish2009, Sharma2010, ZuHone2013, Suzuki2013, Komarov2014, ZuHone2015, Dubois2016, Kannan2017, Yang2016}. 

\section{Initial Conditions and Simulations} \label{sec:ICS}

  \begin{table}
    \caption{Overview of the number of particles, the mass resolution, the gravitational softening lengths and the physical field strength of the three targeted resolution levels.}
    \centering
    \label{tab:one}
    \begin{tabular}{ccccc}
      \hline
      \multicolumn{5}{c}{Particle Numbers}\\\hline
				&			& $1$X &  $10$X & $25$X \\
      Gas particles			&$N_\mathrm{g}$	& $\sim 1 \cdot 10^{6}$ & $\sim 1 \cdot 10^{7}$ & $\sim 3 \cdot 10^{7}$\\
      Dark matter		&$N_\mathrm{dm}$	&  $\sim 1 \cdot 10^{7}$ & $\sim 1 \cdot 10^{7}$ & $\sim 3 \cdot 10^{7}$\\
      \hline

      \hline
      \multicolumn{5}{c}{Mass resolution $[M_{\odot}]$}\\\hline
      Gas particles             &$m_\mathrm{gas}$   & $\sim 1.4 \cdot 10^{8}$ & $\sim 1.4 \cdot 10^{7}$ & $\sim 5.6 \cdot 10^{6}$\\
      Dark matter               &$m_\mathrm{dm}$    & $\sim 6.9 \cdot 10^{8}$ & $\sim 6.9 \cdot 10^{7}$ & $\sim 2.7 \cdot 10^{7}$\\
      \hline
      \multicolumn{5}{c}{Gravitational softening $[kpc]$}\\\hline
      Gas particles		&$\epsilon_\mathrm{gas}$	& 3.0 & 1.4 & 1.0\\
      Dark matter		&$\epsilon_\mathrm{dm}$		& 3.0 & 1.4 & 1.0\\
      \hline
      \multicolumn{5}{c}{Seed field (physical)}\\\hline
      & & $5.02 \cdot 10^{-11}$ & $1.98 \cdot 10^{-10}$ & $3.28 \cdot 10^{-10}$\\
      \hline
    \end{tabular}
  \end{table}

We run a suite of six non-radiative (without cooling and star formation) cosmological zoom-in simulations of a galaxy cluster with a target mass of $M_{200} \sim$ $2 \cdot$ 10$^{15}$ M$_{\odot}$ to the target redshift $z=0$. We run the simulations at three different mass resolutions which we will refer to as $1$X, $10$X and $25$X. In this naming convention the leading number indicates $1$, $10$ or $25$ times the mass resolution that is achieved in the Magneticum high resolution cosmological volume simulations \citep{Hirschmann2014}. We choose this setup for the following reasons. First, this allows us a detailed study of the plasma physics that is acting within the ICM without being polluted by the additional energy input of SNe or active galactic nuclei (AGN). This enables us to investigate the magnetic field amplification in the ICM without the need of re-tuning feedback parameters on the different resolution levels and directly sets the origin of the turbulence injected in the system to be driven by gravitational forces only. Thus we can link any amplification of the magnetic field directly to the gravo-turbulence within the ICM without considering turbulence driven by stellar or AGN feedback. However, we will comment on the caveats of this in section \ref{sec:conclusions} in greater detail. Second, previous studies of the dynamo in the ICM that were obtained with the grid code \textsc{enzo} presented in \cite {Vazza2018} have been carried out with a similar non-radiative setup which allows for a pristine comparison with their results. 

\subsection{Simulation Setup}
The initial conditions for the cluster at hand are selected from a lower resolution volume with a box size of $1.0$ Gpc and a base resolution of $1024^3$ dark matter particles leading to a particle resolution of around $10^8$ M$_{\odot}$. We use a WMAP$7$ cosmology with $\Omega_{0} = 0.24$, $\Omega_{\Lambda} = 0.76$, $\Omega_{baryon} = 0.04$, $h = 0.72$ and $\sigma_{8} = 0.8$. We select the dark matter particles at $z=0$ in cosmological volume at base resolution and trace them back to a resolution dependent initial redshift using the code \textsc{zic} \citep[][]{Tormen1997}. The code allows for arbitrary shapes of the high resolution regions to avoid overhead by oversampling the high resolution regions when it is simply tied to a sphere or an ellipsoid. The initial redshifts for the three resolution levels are as follows: $z_\mathrm{ini}=70$ ($1$X), $z_\mathrm{ini}=140$ ($10$X) and $z_\mathrm{ini}=180$ ($25$X).\\ The region that is selected for re-simulation is chosen to be large enough to avoid the pollution of the target halo at $z=0$ by lower resolution intruder particles that originate from the lower resolution large scale structure in the Gpc volume. For each resolution level one dark matter only test run has been carried out to ensure the quality of the initial conditions in which the gas particles have been co-evolved as separate dark matter species. As the mass resolution and the applied force softening change on every of the three resolution levels we sum up the basic simulation parameters in Table \ref{tab:one}. We note that the force softening in our lowest resolution run is $3.0$ kpc which is still smaller than the spatial resolution of the highest resolution run of \citet{Vazza2018}. In our highest resolution run the force softening is pushed to one kpc and thus the resolution on all resolution levels is sufficient to indicate magnetic field amplification via the small-scale turbulent dynamo. As we carry out all simulations in a non-radiative fashion we need an initial seed field as this is the only possible origin of the magnetic field in a scenario like this. For this we choose a default value of $B_\mathrm{seed}= 10^{-14}$ G (co-moving). As we start our simulations at different redshifts, this corresponds to a variation in the physical seed field of B$_\mathrm{init, ph} = B_\mathrm{seed} \cdot (1+z_\mathrm{init})^2$ for the different resolutions which we sum up in Table \ref{tab:one}. We note that this is a rather conservative choice for the magnetic seed field and other simulation groups often take larger values \citep[e.g.][]{Vazza2018}.\\
Furthermore, we will test different physical settings on our 10X model. For this purpose we need to introduce naming conventions and model variation parameters.

\subsection{Naming conventions and model variations}

The MHD simulations with the above mentioned default settings are referred to as $1$X, $10$X and $25$X. Throughout the paper we will make quite limited use of our hydrodynamics only models. Therefore, whenever we will refer to one of the hydrodynamics only simulations we will explicitly label it as the hydrodynamic realisation of the model $1$X, $10$X or $25$X. The focus of this work is magnetic field amplification in the ICM and the hydrodynamics-only simulations are merely reference simulations used for the comparison with the MHD runs\footnote{The hydrodynamics models are used as a sanity check for the MHD models in terms of mass and size growth of the objects (see section \ref{sec:cosmo_assembly}).}.\\
Furthermore, we will test three important model variations on the $10$X run and introduce new naming conventions for the specifics of the respective run. The three model variations are as follows:
\begin{enumerate}
    \item First, we will investigate the variation of the magnetic seed field to a ten times smaller and a ten times higher value than the value we chose in our default setting (for $10$X this is $1.98 \cdot 10^{-10}$ G.), resulting in $1.98 \cdot 10^{-11}$ and $1.98 \cdot 10^{-9}$ G. These two simulations are labelled as $10$X-low-seed and $10$X-high-seed and will be subject of Appendix \ref{appendix:A}.
    
    \item Second, we will investigate how robust our results are on the choice of the numerical diffusion parameter. Our default setting for the diffusion parameter is $\sim \eta_\mathrm{turb} = 2 \cdot 10^{27}$ cm$^2$ s$^{-1}$ \citep{Bonafede2011}. In order to understand the dependence of the magnetic field growth on the diffusion parameter we will also vary this to a ten times lower value and a ten time higher value of $\eta_\mathrm{turb} = 2 \cdot 10^{26}$ cm$^2$ s$^{-1}$ and $\eta_\mathrm{turb} \approx 2 \cdot 10^{28}$ cm$^2$ s$^{-1}$, respectively. For these simulations we introduce the naming conventions 10X-low-eta and 10X-high-eta. They will be subject of Appendix \ref{appendix:B}.
    
    \item Third, we will carry out one additional run with anisotropic thermal conduction where we directly include the magnetic field structure in our thermal conduction solver via the a biconjugate gradient solver \citep{Arth2014}. All other simulations are carried out with physical, but isotropic conduction which is a potential caveat in the MHD case. However, isotropic conduction is already computationally expensive to solve and takes up roughly $15$ per cent of the computing time. Anisotropic conduction is even more demanding in terms of computational cost and memory imprint of the code. Thus we only carry out one simulation labelled as 10X-ani with the effect of an anisotropic physical conduction, which will be subject of Appendix \ref{appendix:C}.
\end{enumerate}    

We show an overview of all the simulations with the physics variations that we carried out in this work in Table \ref{tab:models}.

  \begin{table*}
    \caption{Overview of the physics variations adopted throughout our different simulations.}
    \centering
    \label{tab:models}
    \begin{tabular}{lcccccc}
      \hline
	  Name		& non-ideal MHD & $\eta$ [cm$^2$ s$^{-1}$] & Thermal Conduction & Anisotropic Thermal Conduction & $\kappa/\kappa_\mathrm{s}$ \\
	  \hline
      $1$X & \checkmark  & $2 \cdot 10^{27}$ & \checkmark & \text{\sffamily X} & $0.05$\\
      $10$X & \checkmark & $2 \cdot 10^{27}$ & \checkmark & \text{\sffamily X} & $0.05$\\
      $25$X & \checkmark & $2 \cdot 10^{27}$ & \checkmark & \text{\sffamily X} & $0.05$\\
      $1$X-NO & \text{\sffamily X} & - & \checkmark & \text{\sffamily X} & $0.05$\\
      $10$X-NO & \text{\sffamily X} & - & \checkmark & \text{\sffamily X} & $0.05$\\
      $25$X-NO & \text{\sffamily X} & - & \checkmark & \text{\sffamily X} & $0.05$\\
      $10$X-low-seed & \checkmark & $2 \cdot 10^{27}$ & \checkmark & \text{\sffamily X} & $0.05$\\
      $10$X-high-seed & \checkmark & $2\cdot 10^{27}$ & \checkmark & \text{\sffamily X} & $0.05$\\
      $10$X-low-eta & \checkmark &$2 \cdot 10^{26}$& \checkmark & \text{\sffamily X} & $0.05$\\
      $10$X-high-eta & \checkmark & $2 \cdot 10^{28}$ & \checkmark & \text{\sffamily X} & $0.05$\\
      $10$X-ani & \checkmark & $2 \cdot 10^{27}$ & \checkmark & \checkmark & $0.05$\\
      \hline

    \end{tabular}
  \end{table*}

\section{Results} \label{sec:results}
In this section we present the results of our simulations. First, we discuss the cosmological assembly of the structure in terms of halo mass and will investigate the general impact of magnetic field on the structure formation process. This is followed by a detailed investigation of the build up of the magnetic field from the initial redshift down to redshift zero. Finally, we will briefly discuss the impact of the divergence cleaning constraint on the structure of the ICM in our simulated galaxies cluster.

\subsection{Cosmological assembly and general cluster properties \label{sec:cosmo_assembly}}

\begin{figure}
    \centering
    \includegraphics[scale=0.32]{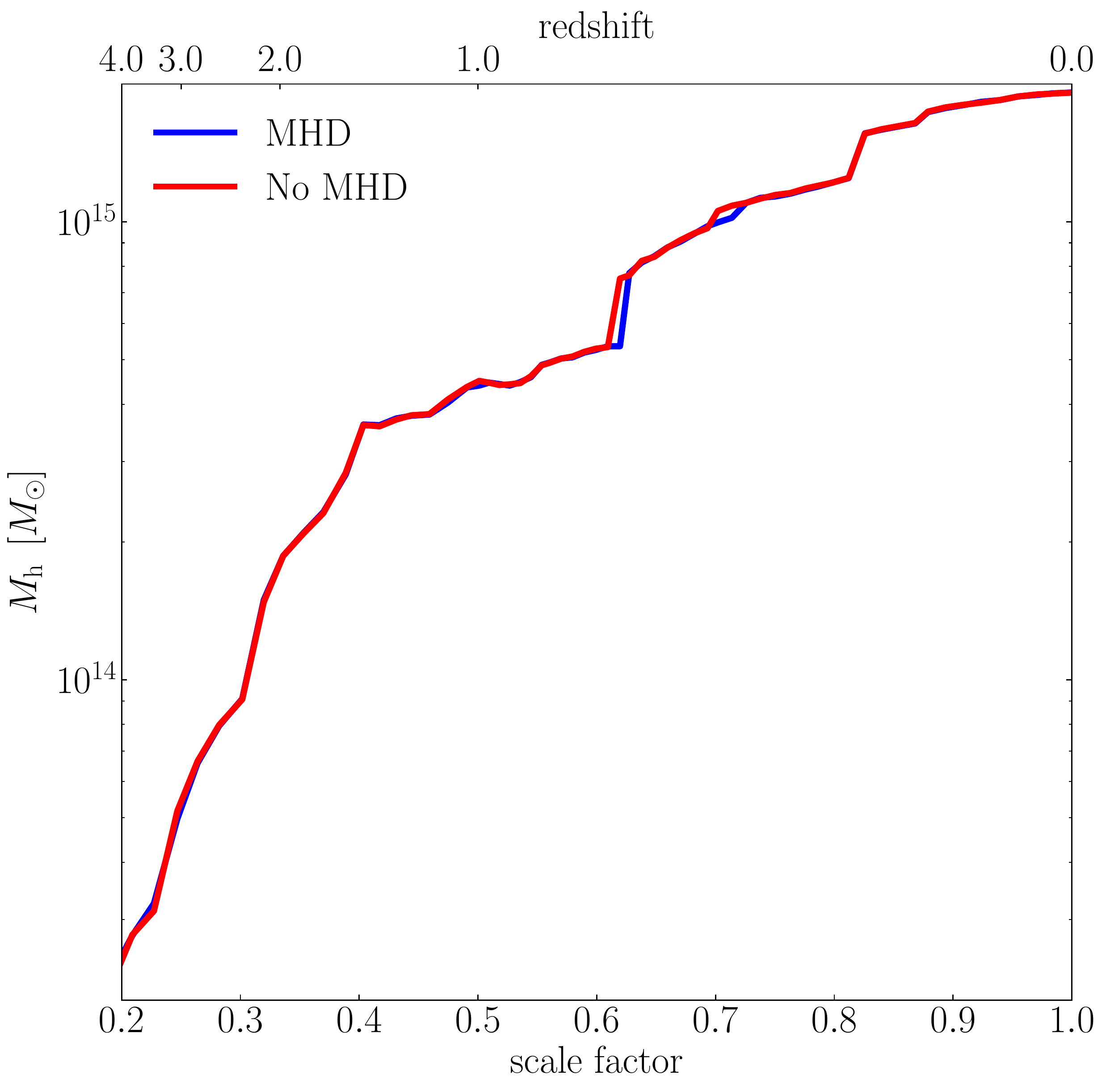}
    \caption{We show the accretion history of the clusters for the MHD case (blue) and for the HD case (red) for reference only for our lowest resolution run to gauge that magnetic fields have a weak effect on the structure formation process.}
    \label{fig:halo_mass_evo}
\end{figure}

First, we briefly discuss the cosmological assembly of our poster-child galaxy cluster to gauge that it reaches a halo mass of around $M_{200} \sim 2 \cdot 10^{15}$ M$_{\odot}$ at redshift zero. We show this in Figure \ref{fig:halo_mass_evo} for our $1$X resolution simulations with and without magnetic field from redshift $4$ to redshift $0$. The structure itself starts to form at a much higher redshift and has already assembled around $\sim 2 \cdot 10^{13}$ M$_{\odot}$ by redshift 4, which is around $1$ per cent of the mass that it will acquire by redshift zero. The cluster undergoes very rapid growth between redshift $4$ and redshift $1.5$ from $\sim 2 \cdot 10^{13}$ M$_{\odot}$ to $\sim 3 \cdot 10^{14}$ M$_{\odot}$ which correspond to a growth rate of around $9.5 \cdot 10^{13}$ M$_{\odot}$ Gyr$^{-1}$. After that the systems transits into a phase of weaker growth until redshift $0.8$ in which it doubles its mass. This is followed by a major merger at redshift $z=0.8$ at which the system roughly acquires another 30 per cent of its total mass up to that point pushing it just below the 10$^{15}$ M$_{\odot}$ mark and is then finally transiting to the regime of continued growth of the system via smooth accretion. This is followed by another major merger at redshift $0.3$. Past redshift $0.2$ the system quietly assembles the rest of its mass until it reaches a final mass of $\sim 2 \cdot 10^{15}$ M$_{\odot}$ at redshift zero. We note that the reference run without the effects of magnetic field (red line in Figure \ref{fig:halo_mass_evo}) is following the MHD run very closely with an error below the $1$ per cent margin for most of the evolution of the system. This gauges the expected very weak effect of the presence of the magnetic field on the large scale assembly of the structure and shows that the magnetic field is not altering the behaviour of the cosmological assembly of the structure. However, there is one exception to this, which is the slight delay of the first major merger of the system at around redshift $0.8$ which slightly delays the merger which could potentially originate from the fact that the additional pressure component that is present as the magnetic field within the system is slowing down the collapse of the baryons into the dark matter halo which indirectly slows down the assembly of the dark matter mass in the centre of the halo.

\subsection{Morphology of the cluster}
\begin{figure*}
    \centering
    \includegraphics[scale=0.35]{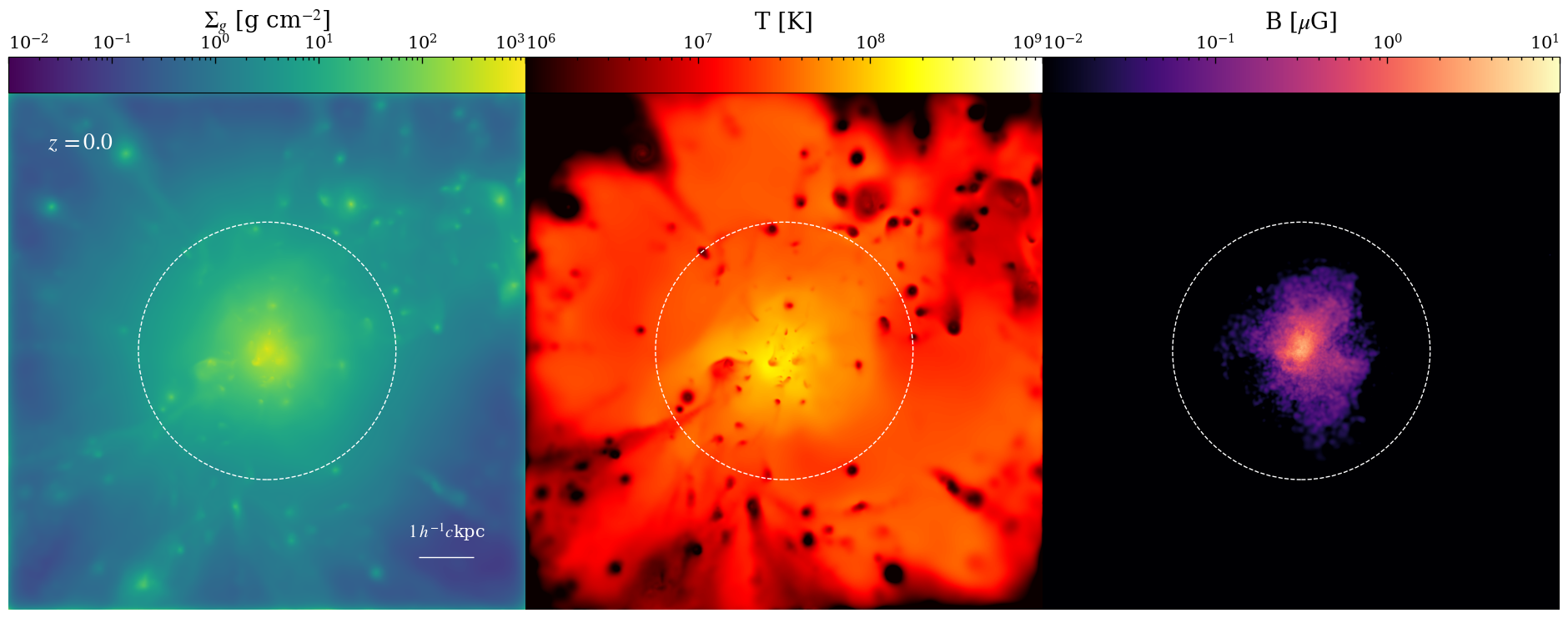}
    \includegraphics[scale=0.35]{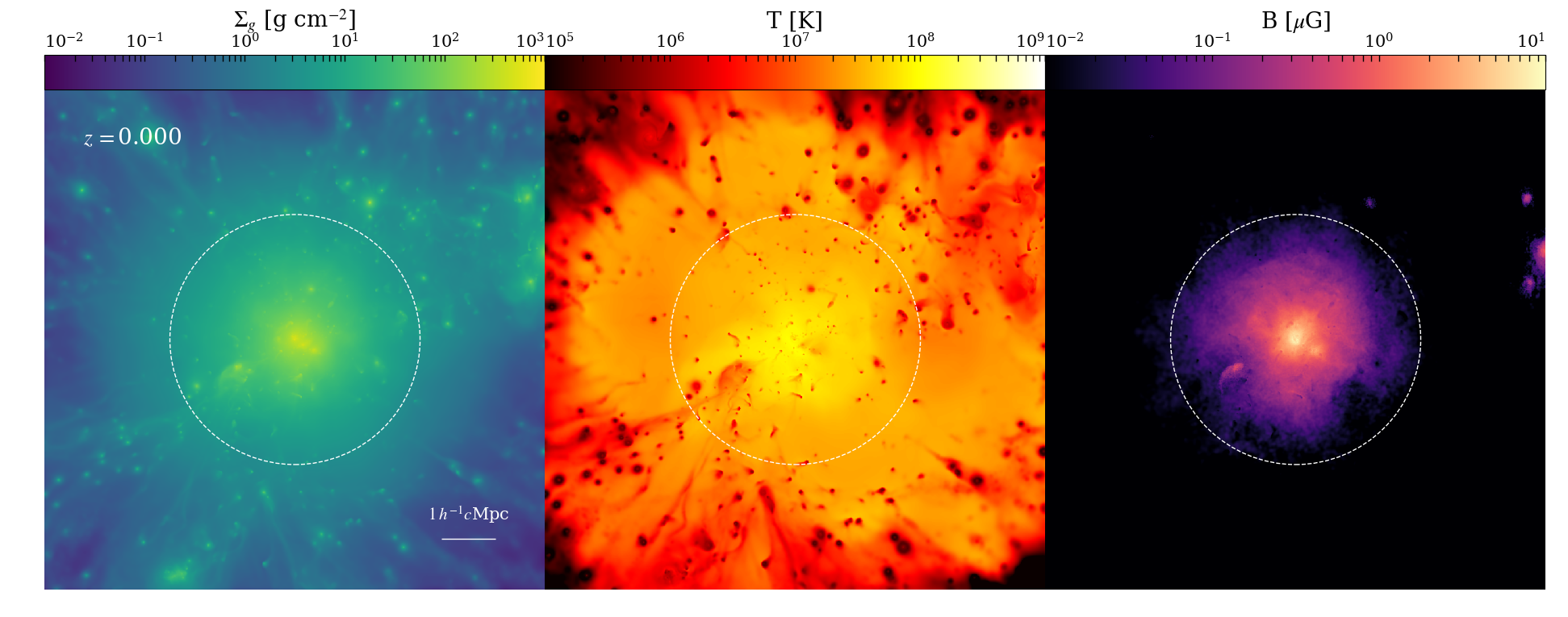}
    \includegraphics[scale=0.35]{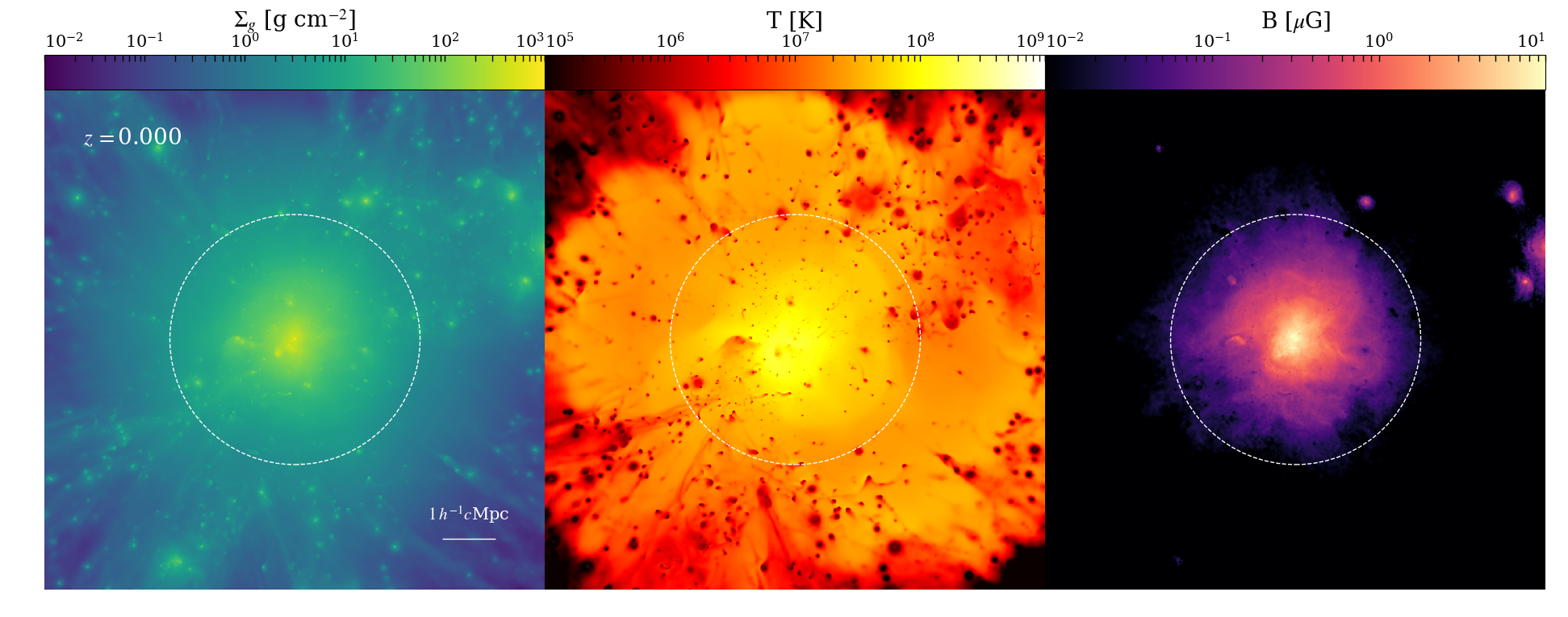}
    \caption{We show the projections of density (left), temperature (centre) and magnetic field (right) for our runs of 1X (top), 10X (centre) and 25X (bottom) at redshift zero. While the 10X and the $25$X simulations show structural similarities, in the magnetic field and temperature structure, there are structural differences to the run 1X. Moreover, the central magnetic field increases by a factor of around $2.5$ from $1$X to $25$X.}
    \label{fig:prop_fig1}
\end{figure*}

\begin{figure*}
    \centering
    \includegraphics[scale=0.35]{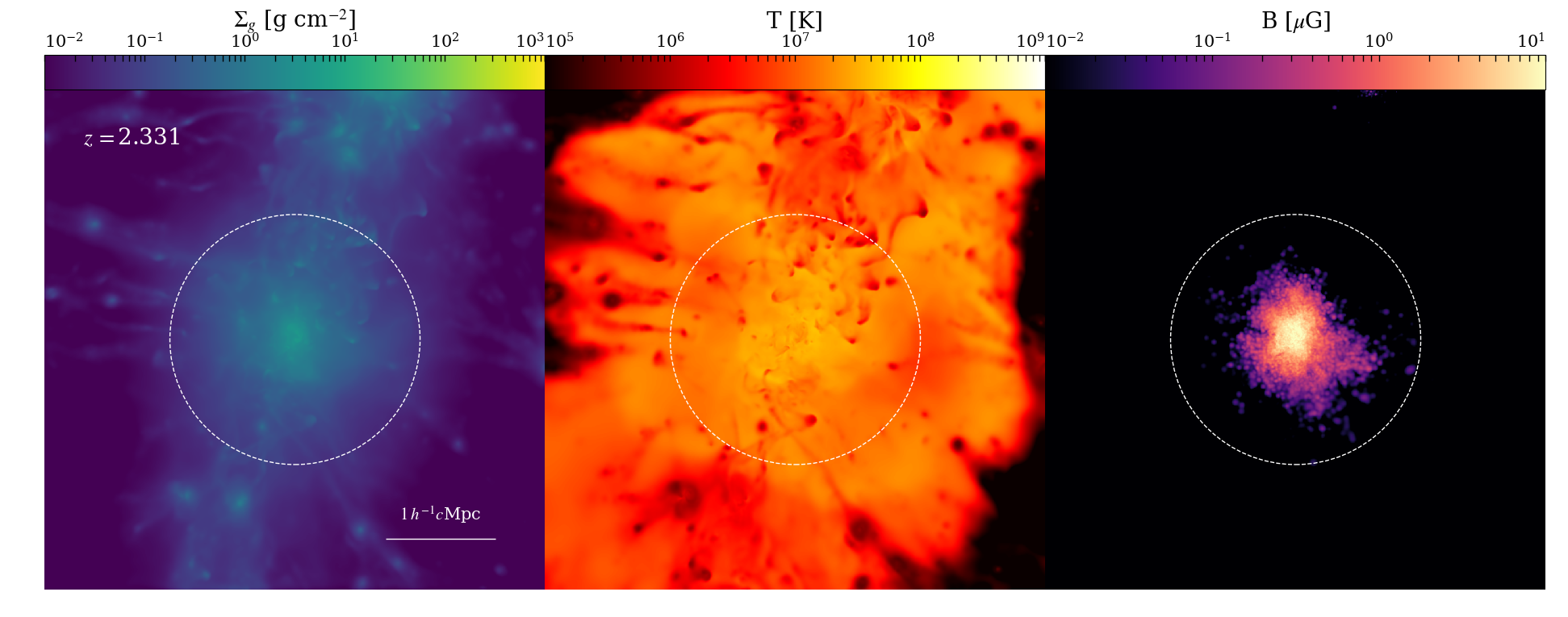}
    \includegraphics[scale=0.35]{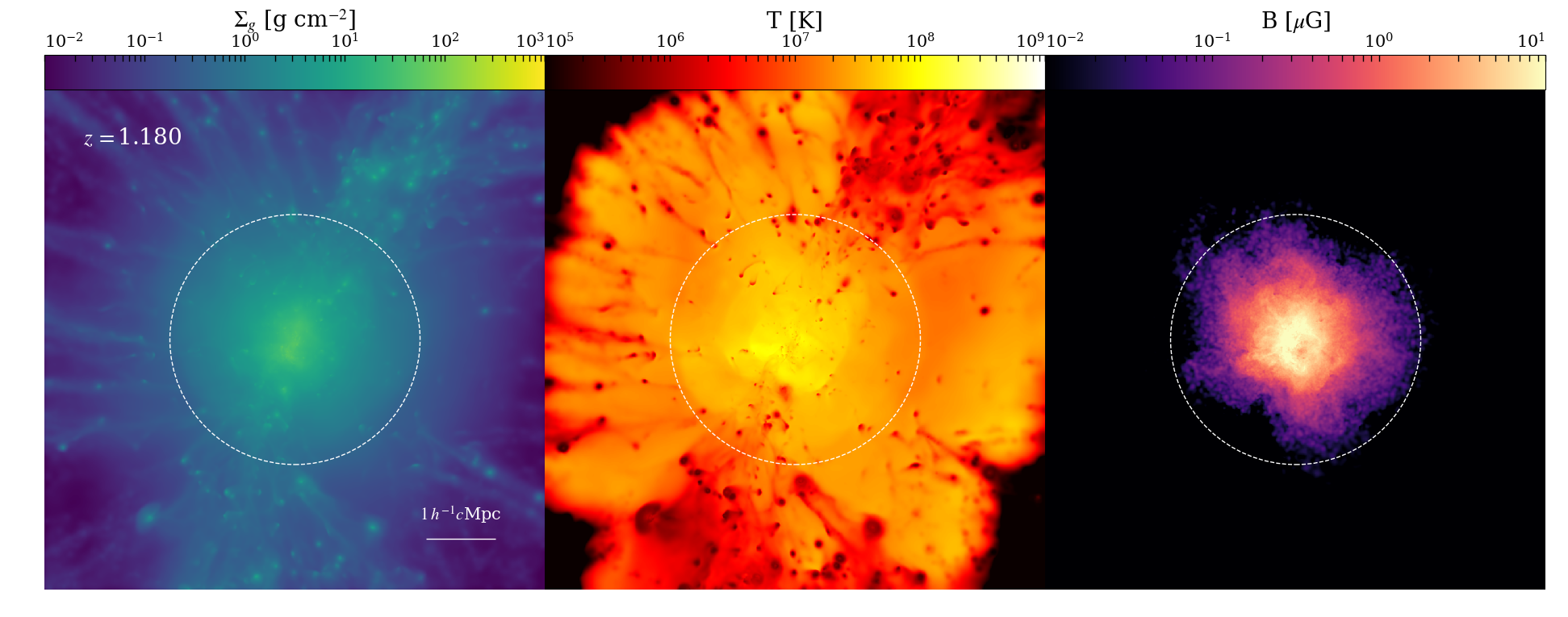}
    \includegraphics[scale=0.35]{plots/MMH_2Rvir_072_25X_Rho_T_B.png}
    \caption{We show the projected density (left), temperature (centre) and magnetic field (right) for three different points in time, at $z\sim 2$ (top), $z\sim 1$ (centre) and $z\sim 0$ (bottom) for our $25$X run. We can clearly see that we have a fully developed magnetic field structure by redshift $z \sim 2$.}
    \label{fig:prop_fig2}
\end{figure*}

We start the evaluation of our results by visualising the key quantities of the cluster for our different runs in Figure \ref{fig:prop_fig1}. In the top row we show the model $1$X, in the middle row we show the model $10$X and in the bottom row we show the model $25$X. The panels on the left show the gas surface density, the panels in the centre show the temperature distribution of the cluster and the panels on the right show the magnetic field strength in the three different models. The white dashed circle in the centre of each panel indicates the virial radius of the cluster.  In the $1$X run we can see a clear lack of resolution, especially in the cluster outskirts beyond the virial radius of the system. This manifests as vanishing substructure in the density distribution compared to the higher resolution models $10$X and $25$X. However, the largest difference between the $1$X model and the $10$X and $25$X models can be seen in the temperature and magnetic field distributions. Visually it appears that there is more hot gas around $10^8$ K in the virial radius for the two higher resolution models $10$X and $25$X. Furthermore, we can identify a clear trend of an increase of the magnetic field strength within the virial radius by a factor of around three from the $1$X model to the $10$X and $25$X models. In this context we want to note that the particles with the maximum field strength within the simulation are located around the cluster centre. For the $1$X simulation the particle with the maximum field strength has a value of $\sim120$ $\mu$G, for the $10$X run we find $\sim 180$ $\mu$G and for the $25$X run we find $\sim 240$ $\mu$G. We note that there are very few particles on each resolution level that have similar magnetic field strength (its around 10 particles for the $25$X simulation that have a field beyond 100 $\mu$G). We will discuss in \ref{sec:divb} to which degree this behaviour is driven by our non-zero divergent in all the simulations. Despite the slightly larger field strength in the runs $10$X and $25$X we want to point out the increase of magnetic field line structures that we can capture in the higher resolution runs $10$X and $25$X compared to the $1$X run. \\
Moreover, we gauge the assembly of the cluster as a function of redshift for the run $25$X for three different redshifts in Figure \ref{fig:prop_fig2}. We show the same quantities as in Figure \ref{fig:prop_fig1}. The top row of Figure \ref{fig:prop_fig2} shows the cluster at $z=2.331$, the second row shows the cluster at a redshift of $z=1.180$ and the third row is the same as the third row of Figure \ref{fig:prop_fig1} showing the cluster at redshift $z=0$. On can see that the central density of the cluster is continuously increasing as a function of redshift which is happening by subsequent merger and accretion events. We can see at redshift $z=2.331$ that a massive structure is about to fall in from the top right. At $z=1.180$ we can see smaller structures falling in from beyond the virial radius, that show extended tails of stripped gas. At redshift $z=0$ the cluster evolves to a more relaxed state with a lower number of in falling objects. There is also a clear evolution in the temperature profiles that we can see in the centre panels from top to bottom where the cluster gas is strongly heated through its formation process down to redshift $z=0$. The magnetic field structure is of particular interest as it is apparent from the evolution of the magnetic field on the right hand side of Figure  \ref{fig:prop_fig2} from top to bottom, that we can find a fully developed magnetic field with around a few $\mu$G already at redshift $z=2.331$ that seems visually to decrease but occupies a larger volume as the system evolves towards redshift $z=1.180$. At redshift $z=0$ we find a fully developed field within the virial radius. Visually, the magnetic field amplification seems to be correlated with the turbulence that is injected via the structure formation process. We want to specifically point out that the field is stronger at around redshift $z=2$ and $z=1$ compared to the field strength at redshift $z=0$.

\subsection{Radial evolution of the cluster}
\label{sec:radial}
\begin{figure*}
    \centering
    \includegraphics[scale=0.6]{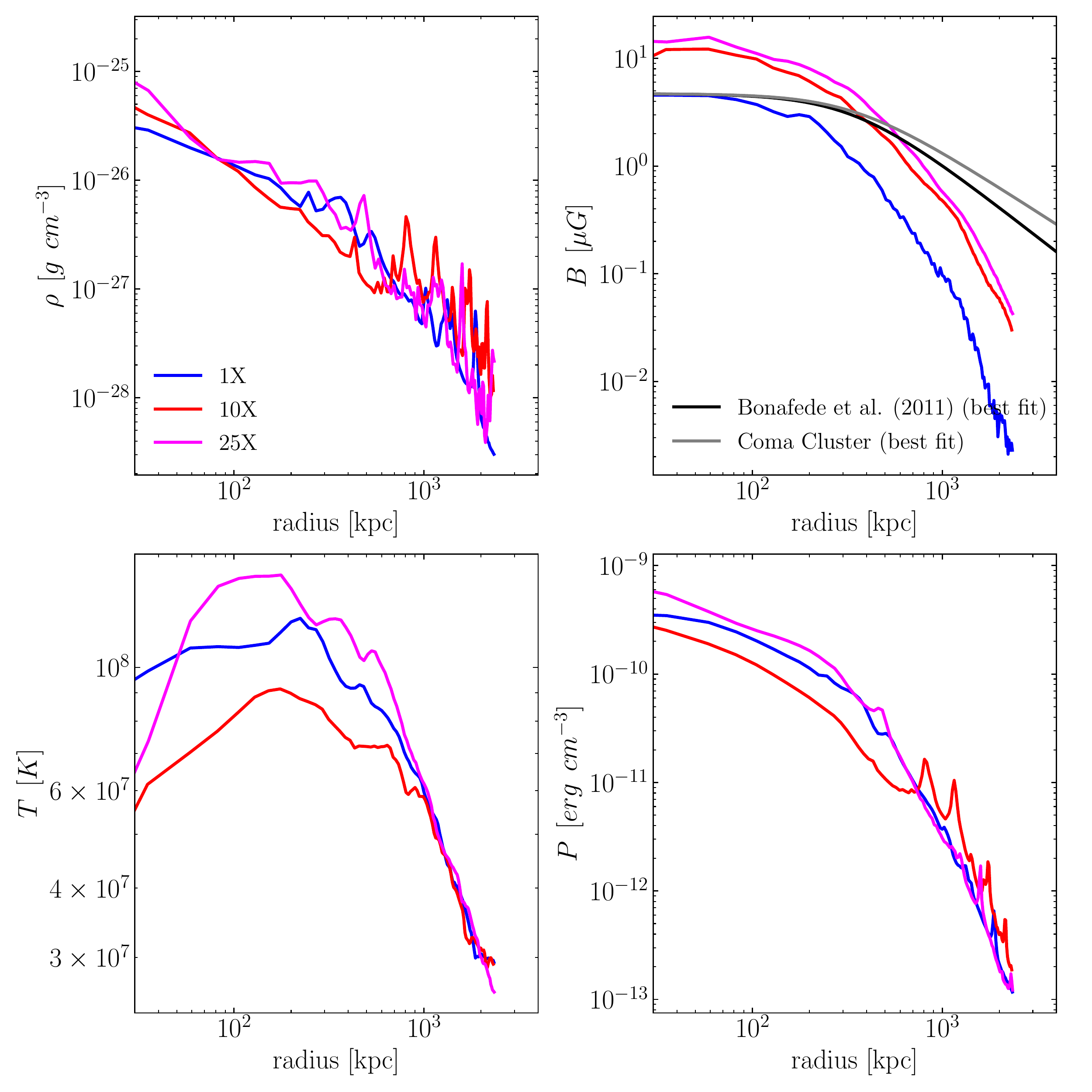}
    \caption{We show radial profiles of the density (top left), the volume weighted magnetic field strength (top right), the temperature (bottom left) and the thermal pressure (bottom right) for our three galaxy cluster simulations $1$X (blue), $10$X (red) and $25$X (magenta). For the magnetic field we over plot the best fit from \citet{Bonafede2011} (black line) and observations of the magnetic field within the Coma galaxy cluster from \citet[][]{Bonafede2010} (grey line). 
    }
    \label{fig:radial_z0}
\end{figure*}

Before we start the discussion on magnetic field amplification via the turbulent dynamo in the ICM we want to briefly report on our results for radial profiles of central physical quantities at redshift $z=0$. In Figure \ref{fig:radial_z0} we show the radial profiles out to a radius of $4$ Mpc for the density (top left), the magnetic field (top right), the temperature (bottom left) and the pressure (bottom right) for the runs $1$X (blue), $10$X (red) and $25$X (magenta). For all quantities we find declining profiles as the function of the radius. As we are specifically interested in the magnetic field evolution of the cluster we note the most important findings regarding the radial trend of the magnetic field strength as a function of resolution. As the resolution is increasing from $1$X to $10$X and finally to $25$X we find an increase of the central magnetic field from around $5$ $\mu$G in the case of the $1$X simulation over $9$ $\mu$G in the $10$X simulation to $14$ $\mu$G in the $25$X simulation. We compare our predicted magnetic field profiles from our simulations to the best fit to a $\beta$-model from the observations of the magnetic field in the Coma galaxy cluster (grey line in the top left panel of Figure \ref{fig:radial_z0}) and the best fit obtained from $24$ simulations of galaxy clusters at the same resolution then our $1X$ run from the same parent dark matter box from \citet{Bonafede2011} (black line in the top panel of Figure \ref{fig:radial_z0}). While our results for the $1$X run are in good agreement with respect to the central magnetic field value in the cluster compared to Coma observations and the simulations of \citet{Bonafede2011} our higher resolution models over predict the central magnetic field value roughly by a factor of $2.5$. We will investigate the origin of this behaviour in greater detail in Appendix \ref{appendix:A} and Appendix \ref{appendix:B} by varying the magnetic diffusion constant and the initial seed field strength. Despite the fact that our higher resolution simulations predict a central magnetic field strength that is higher than observed values in the coma cluster we note that state-of-the-art simulations with Eulerian gird codes typically predict values that just reach the $\mu$G regime and are a around the same factor too low compared to observed values within the coma cluster that report central field strengths of around 7 $\mu$G \citep{Bonafede2010}. Moreover, we note that the cluster that we simulated is not really a Coma-cluster analog as this is a system that is in equilibrium at redshift $z=0$ and the coma cluster is not \citep[e.g.][]{Lyskova2019}. Furthermore, as noted above we do not vary the parameters for seed field and diffusion constant which might impact the radial magnetic field distribution at redshift zero. We chose to do this in our default simulation runs to obtain pristine conditions for our study on the galactic dynamo, which is the central subject of this paper. Last but not least the too high central field could also be related to our non-vanishing divergence of the field. We will discuss the impact of the divergence cleaning constraint on the too high central magnetic field strengths in section \ref{sec:divb}.

\subsection{Amplification of the magnetic field}
\label{sec:dynamo}
\begin{figure}
    \centering
    \includegraphics[scale=0.32]{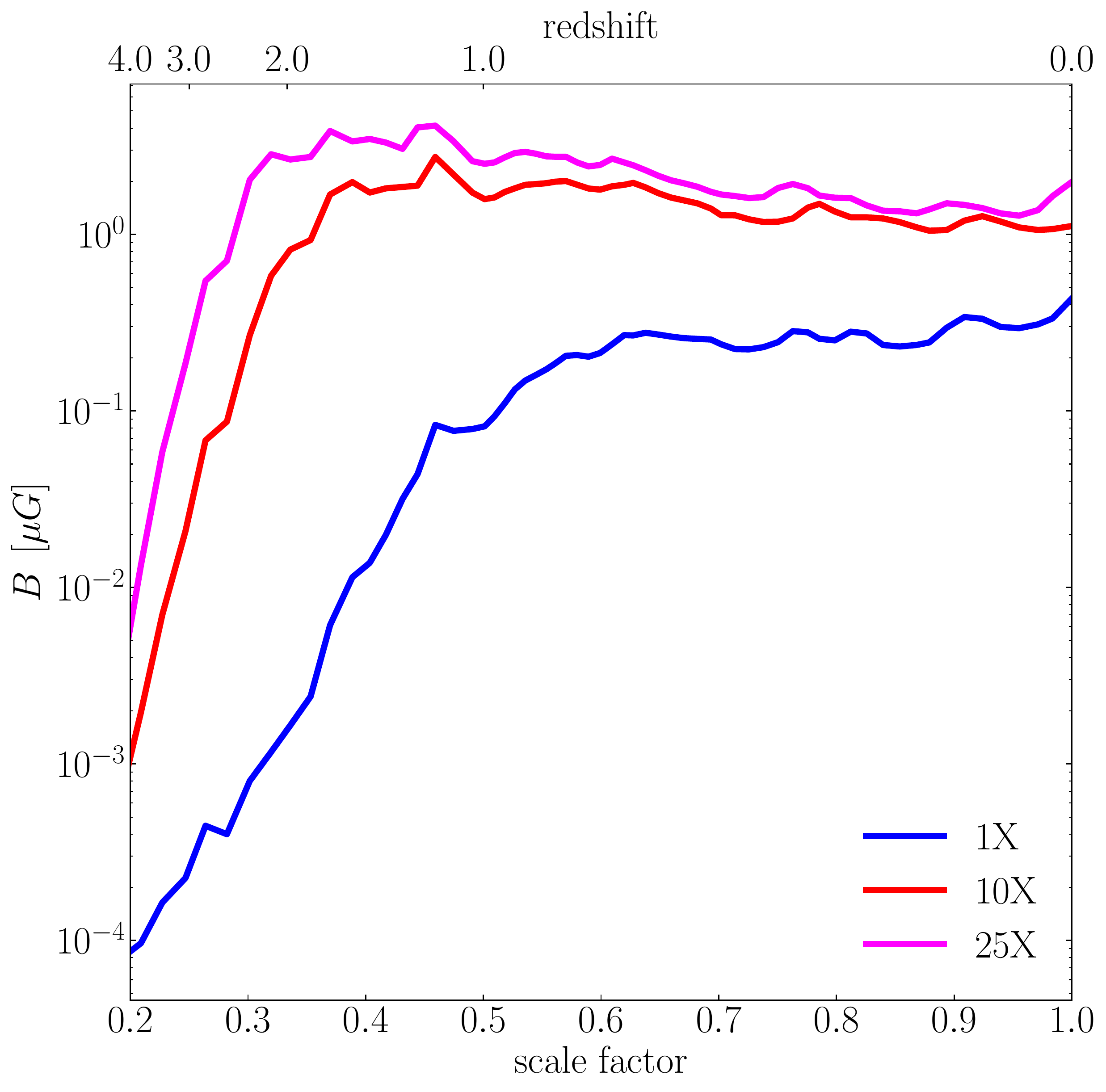}
    \caption{We show the time evolution of the total magnitude of the magnetic field as a function of redshift. This indicates exponential growth of the magnetic field early in the formation history of our galaxy cluster simulations. We indicate the different resolution levels with red (1X), blue (10X) and magenta ($25$X). Bottom: We show the evolution of the magnetic field energy for our cluster for all resolution levels and find only a very weak dependence of the resolution.}
    \label{fig:bfld_evo}
\end{figure}
\begin{figure*}
    \centering
    \includegraphics[scale=0.45]{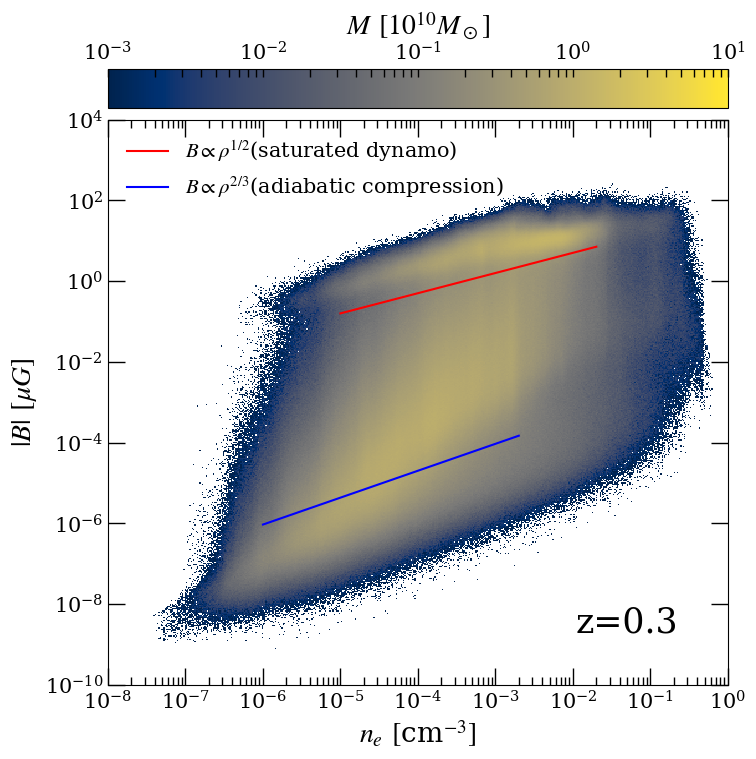}
    \includegraphics[scale=0.45]{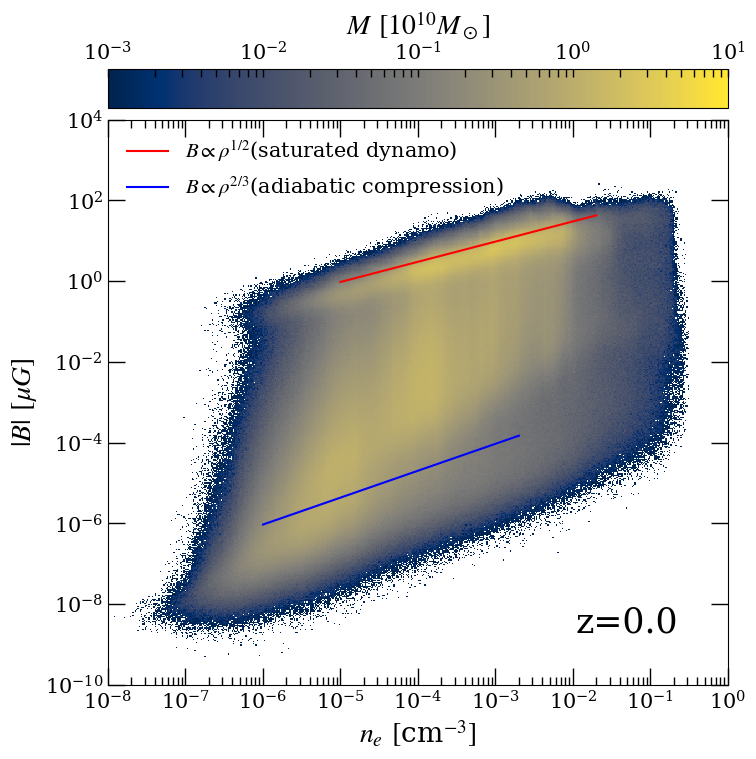}
    \caption{We show the evolution of the density-magnetic field strength phase-space at redshift $0.3$ (left) at which the system undergoes a heavy major merger and at redshift $0$ at which the system transits towards a relaxed state. We include all the gas within R$_\mathrm{vir}$. In both cases we find excellent agreement with the adiabatic compression limit ($B \propto \rho^{2/3}$, blue line) at low magnetic field strengths and lower densities. This can be associated as the gas that is falling towards the centre of the structure. However, in the regime of higher magnetic fields and higher densities we find good agreement with the power-law scaling that is expected from a saturated turbulent dynamo ($B \propto \rho^{1/2}$) at redshift $0$, while there appears to be some deviation from the saturated dynamo at redshift $0.3$, which could hint towards the non-linear dynamo regime.}
    \label{fig:blfd_phase_space}
\end{figure*}
\begin{figure}
    \centering
    \includegraphics[scale=0.32]{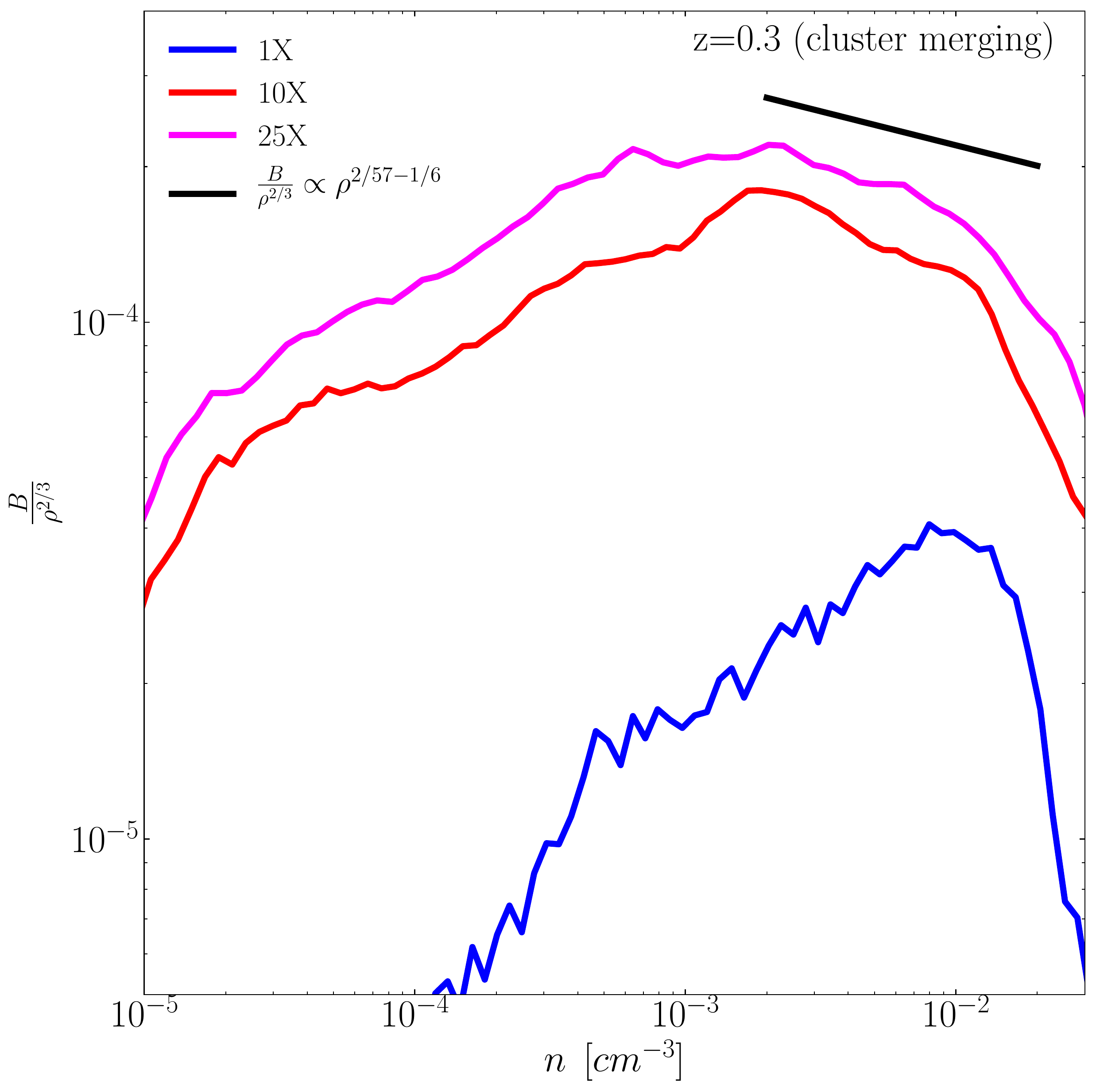}
    \includegraphics[scale=0.32]{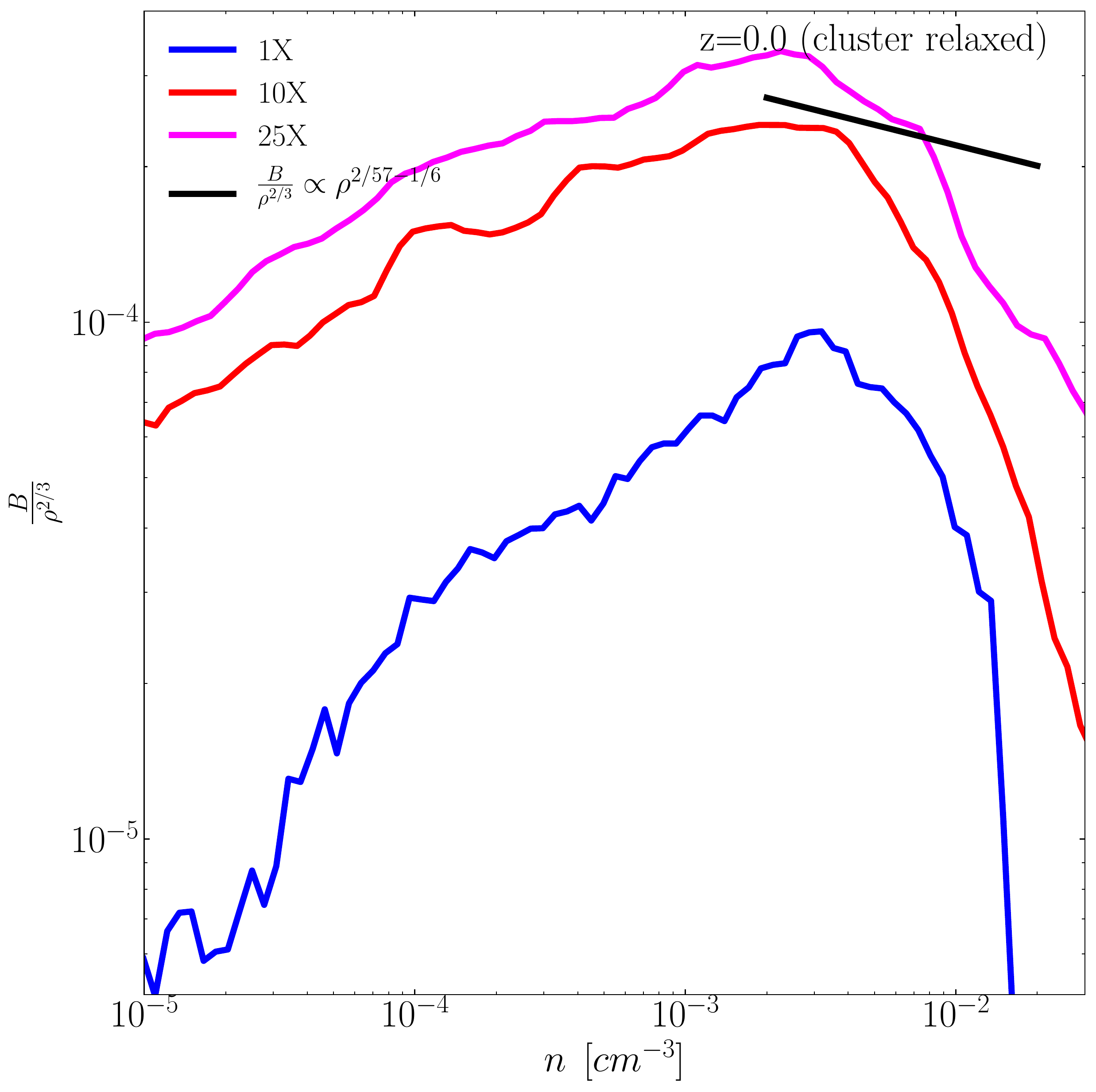}
    \caption{We show the quantity $B/\rho^{2/3}$ as a function of the density in equal log bins to show the agreement of the theory of the turbulent dynamo under gravitational collapse derived by \citet{Xu2020} with our cosmological galaxy cluster simulations at redshift $z=0.3$ at which the cluster is under going gravitational collapse after a major merger(top). At redshift $z=0$ when the cluster is dynamically relaxed we find deviations from the above scaling. The colours indicate our $1$X (blue), $10$X (red) and $25$X (magenta) runs.}
    \label{fig:xu_lazarian}
\end{figure}
\begin{figure}
    \centering
    \includegraphics[scale=0.32]{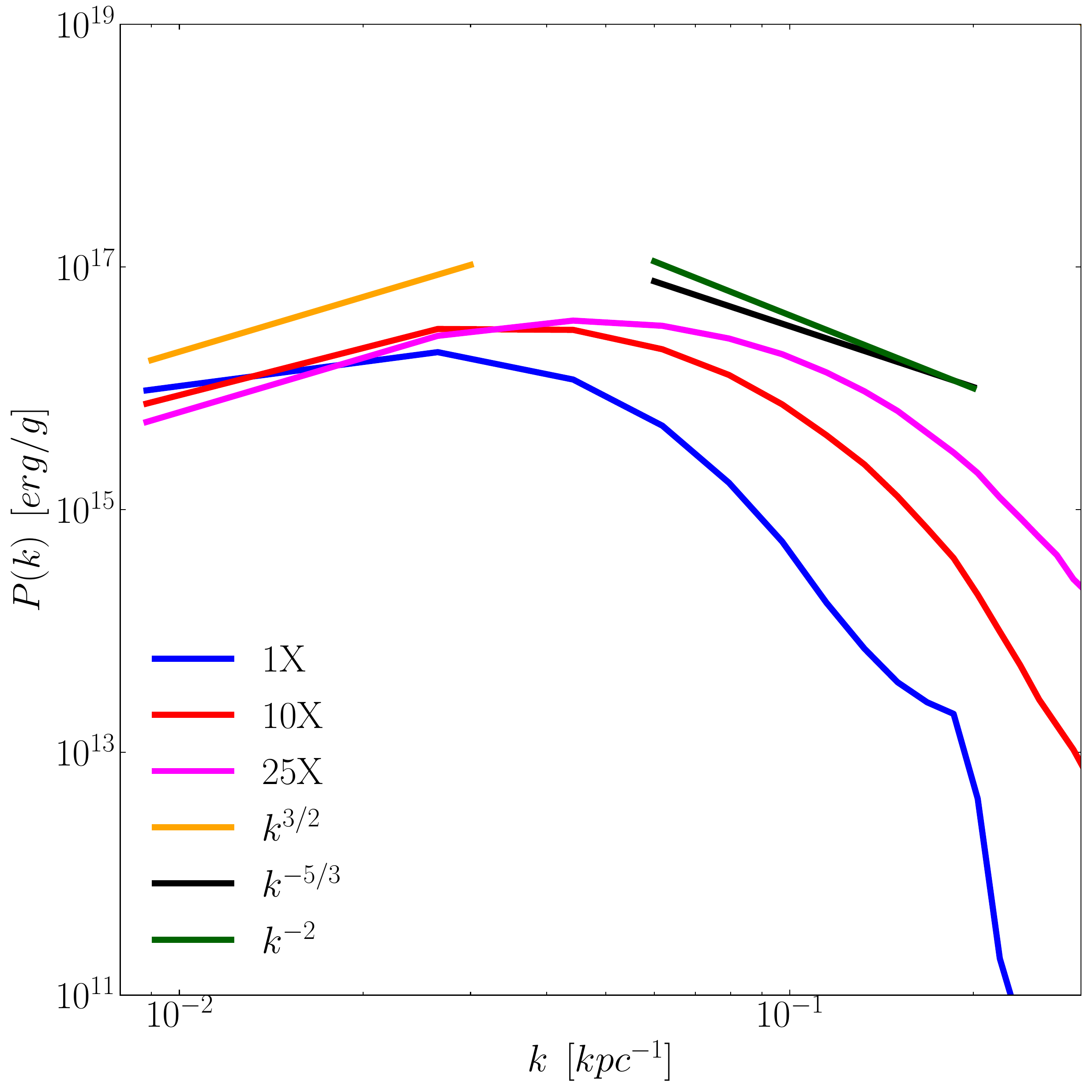}
    \caption{We show the magnetic power-spectra at redshift $z=0$ for all three resolution levels in blue ($1$X), red ($10$X) and magenta ($25$X) of our galaxy cluster zoom simulations. We see very good agreement with the predicted slope from the dynamo theory by \citet{Kazantsev1968} on the large scales (golden line) in our $10$X and $25$X resolution runs. However, we note that the $1$X model is predicting a slightly steeper slope than the $k^{3/2}$ slope from \citet{Kazantsev1968}, which is supposedly related to a lack of resolution. Furthermore, we overplot the k$^{-5/3}$ and k$^{-2}$ slope for reference on the smaller scales. We find a stepper slope on smaller scales than predicted by these scalings which is in accordance with small-scale simulations of the turbulent dynamo \citep[see e.g.][]{Schekochihin2004, Porter2015}.}
    \label{fig:power_z0}
\end{figure}
\begin{figure}
    \centering
    \includegraphics[scale=0.32]{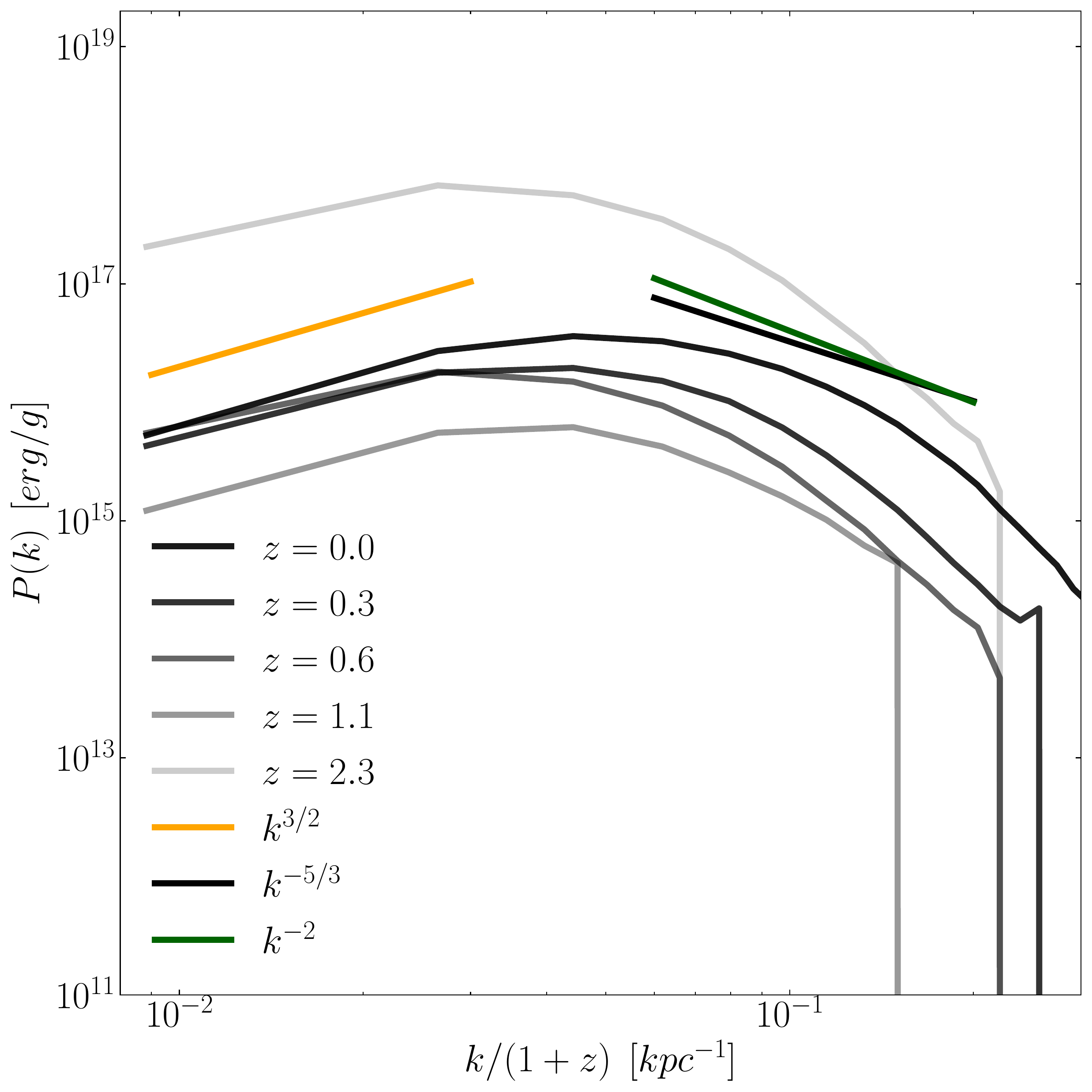}
    \caption{We show the magnetic power spectra for 4 different redshifts of our highest resolution run $25$X. For all presented redshifts we find very good agreement with the predicted slope from \citet{Kazantsev1968} on larger scales (golden line). We overplot the k$^{-5/3}$ and k$^{-2}$ slope for reference on the smaller scales.} 
    \label{fig:power_z_range}
\end{figure}
\begin{figure}
    \centering
    \includegraphics[scale=0.32]{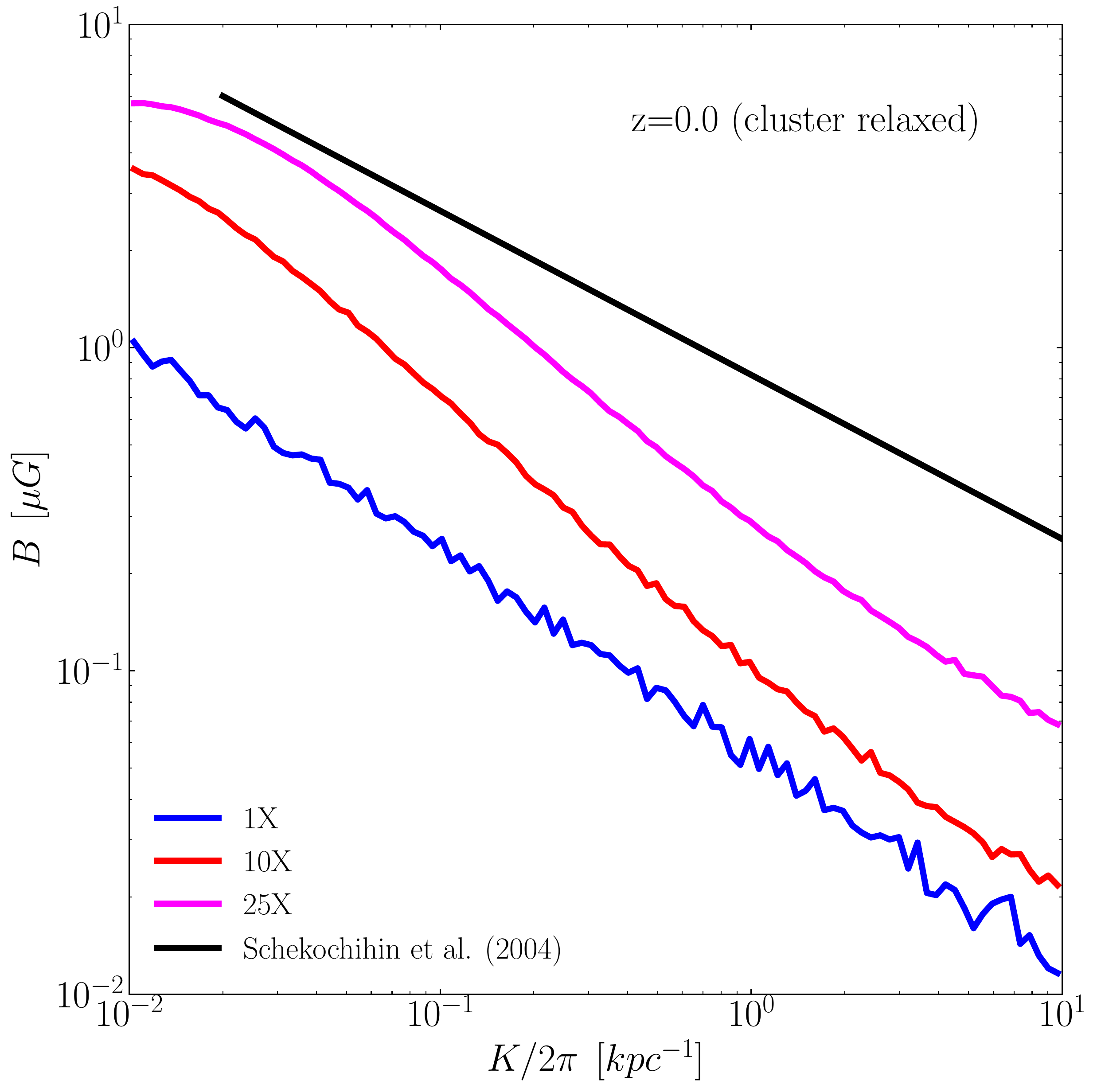}
    \caption{We show the relation between magnetic field strength and magnetic field line curvature for all of our simulation runs $1$X (blue), $10$X (red) and $25$X (magenta). We compare our results to the scaling derived from high resolution idealised dynamo simulations of \citet{Schekochihin2004}. We can recover the decreasing trend following $KB^{1/2} = \mathrm{constant}$ (black line). However, we find  a tilt and a slightly steeper slope then expected in our higher resolutions simulations for the models $10$X and $25$X.}
    \label{fig:curvature_plot}
\end{figure}
We start the discussion about magnetic field amplification in our galaxy cluster zoom-in simulations by considering the time evolution of the magnetic field within one virial radius (R$_\mathrm{vir}$) from redshift $4$ to redshift $0$. We show this in Figure \ref{fig:bfld_evo}. The magnetic field increases exponentially from the initial seed field value between redshift $4$ and redshift $2$ to a sub-equipartition value of around $0.05$ $\mu$G in the $1$X simulation. In the higher resolution simulations $10$X and $25$X we find a very different shape of the growth of the field as a function redshift. Here, the magnetic field in the cluster peaks at around redshift $z=2$ at a value of a few $\mu$G. From that point in time the field decreases towards redshift zero and settles at around $1$ to $2$ $\mu$G within the virial radius. This behaviour is consistent with other cosmological simulations of magnetic field amplification \citep[see e.g.][]{Garaldi2020}. While the exponential increase of the magnetic field strength could potentially be related to magnetic field amplification by a small-scale turbulent dynamo driven by sub-sonic turbulence in the ICM it is impossible to determine this from the evolution of the magnetic field alone. However, we can still estimate the growth rate of the magnetic field for the different runs. Essentially, we find that all three models are initially consistent with exponential growth of the form:
\begin{align}
  B \propto \exp(\gamma \cdot t).
\end{align}
For the $1$X run we find that $\gamma \approx 0.7$ Gyr$^{-1}$ while for the $10$X and $25$X run we find $\gamma \approx 0.15$ Gyr$^{-1}$. The former growth-rate is indicating unresolved dynamo action in the $1$X simulation. Furthermore, we note that the increase of the field we observe towards higher redshift is roughly consistent with an increase of the magnetic field toward higher redshift following the relation:
\begin{align}
    B = B_{z=0} (1+z)^m,
\end{align}
with a power law index of around $m=0.5$. This is in relatively good agreement with the predictions made for \textsc{ska} by \citet{Krause2009}.
It is intrinsically complicated to identify dynamo action in numerical simulations of galaxy and galaxy cluster formation. This is mainly due to the fact that the fundamentals of dynamo theory are built on top of the theory of turbulence, which is generally not very well understood in hydro dynamical numerical simulations. Generally, dynamos work by converting (turbulent) kinetic energy into magnetic field energy on the scale of small turbulent eddies. This process is saturated once equipartition between turbulent kinetic energy and magnetic field energy is reached. The magnetic field energy can then be transported to the larger scales in an so called inverse turbulent cascade. However, the amplification of tiny magnetic seed fields by turbulence is competing with the dissipation of magnetic field on the smallest scales. Only if the interplay between dissipation of magnetic field and transport of the magnetic field alongside its amplification is modelled correctly the dynamo will transit from the linear growth regime, into the non-linear regime and finally saturate. The crux in achieving this is to have enough resolution on small scales to capture magnetic field amplification by turbulence but also enough resolution on the larger scales to model magnetic field transport towards larger structures. This has been subject of MHD research in many Eulerian grid codes in recent years \citep[e.g.][]{Ryu2008, Beresnyak2016, Schekochihin2004, Cho2009, Porter2015, Vazza2018} but there is little to no work on magnetic field amplification in Lagrangian methods. This paper is explicitly targeted to close the gap between the state of research in studies of magnetic field amplification within the ICM that has been put forward in recent years with Eulerian codes. In the following we will present evidence for an acting small-scale turbulent dynamo in the ICM of our simulated galaxy clusters and evaluate the resolution dependence of the process by directly comparing to the dynamo theory that has been put forward by \citet{Kraichnan1967} and \citet{Kazantsev1968} and has been refined by several authors since then \citep[e.g.][]{Zeldovich1983, Kazantsev1985, Kulsrud1992, Kulsrud1997, Subramanian2002, Xu2020}.\\
First, one can study the structure of the cluster in the density-magnetic field phase-space to gauge the dependence of magnetic field on its environment. We show this in Figure \ref{fig:blfd_phase_space} for our highest resolution simulation ($25$X) and two different redshifts, redshift $z=0.3$ (left) and redshift $z=0$ (right). The gas cells are selected within one R$_\mathrm{vir}$ around the centre of the cluster which has been identified with the \textsc{subfind} algorithm \citep{Springel2001, Subfind2009}. The colour code indicates the cell mass and shows how much mass is contained in each state in $10^{10}$ M$_{\odot}$. We deliberately choose these two points in time to distinguish between the linear and non-linear dynamo regime at redshift $z=0.0$ and $z=0.3$ respectively. Within this time frame the cluster transits from a turbulent merging epoch towards a dynamically relaxed system. At both redshifts we can identify gas at low magnetic field strengths and lower gas densities that is in good agreement with the power-law scaling obtained from the flux-freezing regime of ideal MHD of an adiabatically collapsing system ($B \propto \rho^{\frac{2}{3}}$). At larger densities and higher magnetic fields we can identify a different scaling, specifically at redshift zero. We find excellent agreement with the saturated dynamo regime with the scaling $B \propto \rho^{\frac{1}{2}}$ in the framework of reconnection diffusion \citep[see e.g.][]{Xu2020}. While the saturation regime is in very good agreement with our redshift $z=0$ results, this is not the case at $z=0.3$ when the system undergoes a merger event followed by rapid smooth accretion of gas mass towards the cluster centre. While some gas at lower densities is still following the saturation regime there is a clear deviation in the high density tail, that is identifiable as flattening of the scaling followed by a kink within the distribution at a density around n$_{e} = 3 \cdot 10^{-3}$ cm$^{-3}$. This could be evidence for rapid diffusion of magnetic field at the highest densities which would be in agreement with the recent theory proposed by \citet{Xu2020} who study the turbulent dynamo in the framework of reconnection diffusion under gravo-turbulence in cooling star forming cores. While \citet{Xu2020} point out that the theory they develop could be of paramount importance in the regime of first star formation, it is apparent that the idea of gravo-turbulence is of importance on galaxy cluster scales as well. Thus, first we want to point out that the physical systems of a gravitational collapsing star forming core is quite different from the cosmological assembly of a galaxy cluster. In a star forming core the idea would be that cooling is enhancing the collapse as the heat generated by the collapse can efficiently be radiated away. This means, that the system is heavily driven out of equilibrium. However, galaxy clusters are (to first order) virialised and thus in equilibrium, especially if one considers non-radiative simulations of clusters. There is one exception to this, which is when the cluster is undergoing a merger process and the merger remnant continues to accrete material which indirectly mimics the situation for which \citet{Xu2020} derive their dynamo model. \citet{Xu2020} derive the following scaling for the non-linear growth regime of the dynamo under gravitational collapse:
\begin{align}
    \frac{B}{\rho^{\frac{2}{3}}} \propto \rho^{\frac{2}{57} - \frac{1}{6}}. 
\end{align}
\citet{Xu2020} compare their derived scaling to results of simulations of magnetic field amplification of first star formation taken from \citet{Sur2012} and find good agreement of their scaling relation with collapsing star forming structures. However, this is a scale-free problem and can easily be tested in the framework of our cosmological galaxy-cluster simulations. \\
\citet{Sur2012} point out that dynamo amplification under gravitational collapse can better be quantified by evaluating $B/\rho^{2/3}$ than just by evaluating the phase space of magnetic field strength and density in star formation simulations (they simulate the gravitational collapse of a Bonnor-Ebert-sphere). However, the physics that is driving dynamo amplification in the regime of star formation is quite similar to the formation scenario of a galaxy cluster as it undergoes collapse in the dark matter potential and one can directly test the linear growth regime and the non-linear growth regime discussed in \citet{Sur2012} and\citet{Xu2020} respectively in the fashion that is suggested by \citet{Sur2012} in the regime of the formation of a massive galaxy cluster.\\
This is evaluated in Figure \ref{fig:xu_lazarian} where we show the average of $B/\rho^{2/3}$ as a function of the density in equal log bins for our $1$X (blue), $10$X (red) and $25$X (magenta) for redshift $z=0.3$ (top) and $z=0$ (bottom). In this context we can identify the linear growth regime as the monotonically increasing part for as a function of density. For the $1$X (lowest resolution) we find a steeply decreasing part at high densities indicating an unresolved non-linear growth regime at both redshifts of interest. While we find a similar situation for redshift $z=0$ in our two high resolution simulations 1$0$X and $25$X this is different at a higher redshift of $z=0.3$ where we can clearly identify the non-linear growth-regime of the dynamo following the scaling of \citet{Xu2020}. We note that the disagreement with \citet{Xu2020} at redshift $z=0$ can be explained by the fact that our cluster at hand is a dynamically relaxed system at that time. At $z=0.3$ however, the systems is strongly collapsing following a previous major merger, providing the ideal conditions for magnetic field amplification via the refined theory of \cite{Xu2020}. However, despite the agreement with \citet{Xu2020} we already saw the indication for this behaviour in the left panel of Figure \ref{fig:blfd_phase_space} as the kink in the phase-space distribution at a density of roughly n$_{e}3 \cdot 10^{-3}$ cm$^{-3}$ which one can interpret as dissipation of magnetic field in high density regimes.\\
Apart from the density-magnetic field strength phase space there is another way of identifying the small-scale-turbulent dynamo by evaluating the magnetic power-spectra of the simulations. This has become a standard test for identifying an acting small-scale turbulent dynamo in numerical simulations, specifically in the ISM \citep[e.g.][]{Balsara2004, Schekochihin2004, Porter2015, Hennebelle2014, Gent2021} but has recently also become quite popular on the scales of galaxies \citep[e.g.][]{Butsky2017, Martin2018, Martin2020, Pakmor2017, Rieder2016, Rieder2017a, Rieder2017b, Steinwandel2019, Steinwandel2020} and galaxy clusters \citep[e.g.][]{Dubois2008, Ryu2008, Vazza2018} for studying the turbulent dynamo. We show the magnetic power-spectra for our three MHD simulations in Figure \ref{fig:power_z0} for $1$X in blue, for $10$X in red and for $25$X in magenta. We note that these are simply the redshift $z=0$ power-spectra. For each simulation we calculated the power-spectra by binning the SPH-data to a grid. The grid has a resolution of $128^3$ and is represented by a cube with a side length of $3 \cdot R_\mathrm{vir}$. The power on each scale is then computed by evaluating the Fourier modes on the grid scale. \\
One can clearly see that there is excellent agreement of the power-spectra determined by this methodology and the power-law slopes predicted by \citet{Kazantsev1968} for the simulations $10$X and $25$X. For the simulation $1$X we find a slope that is too steep on the large scales to be in good agreement with the \citet{Kazantsev1968} theory which is probably introduced by the lack of resolution we have in the outer parts of the cluster in $1$X compared to $10$X and $25$X. Therefore, we suggest that studies regarding the dynamo on the scales of the ICM require a cell mass resolution of around $10^{7}$ M$_{\odot}$ which corresponds to a spatial resolution of around $2$ kpc. While we generally find little difference between our $10$X and $25$X models we would advise future dynamo studies with Lagrangian methods to adopt our $25$X resolution for converged results on the power-spectra which results in a mass resolution of around $4 \cdot 10^{6}$ and a spatial resolution of roughly $3$ kpc. This is roughly in line with the findings of \citet{Vazza2018} for the grid code \textsc{enzo} who obtain self-consistent power-spectra for their two highest resolution runs. \\
Furthermore, we investigate the power-spectra at four earlier times then redshift $z=0$ for our $25$X run. We show the results in Figure \ref{fig:power_z_range} for $z=0$ (black line), $z=0.3$ (light black line), $z=0.6$ (grey line), $z=1.1$ (light grey line) and $z=2.3$ (very light grey line). Essentially, we find similar results as in the redshift $z=0$ case with the difference that there is less power stored in the magnetic field at higher redshift. Despite the lower power in the magnetic field we are still able to recover a power-spectrum that is in accordance with the theory of \citet{Kazantsev1968} already at redshift $z \sim 1$, after half of the cosmological evolution of the cluster itself. However, we note an exception to this at redshift $z=2.3$ where we find more more power in the magnetic field compared to the lower redshift spectra. This is roughly consistent with he peak in the time evolution of the magnetic field strength at around redshift $z=2$ and generally also consistent with expectations for higher redshift clusters \citep[e.g.][]{Krause2009}.\\
Most studies on the turbulent dynamo on galaxy and galaxy cluster scales stop at the point where they achieve the scalings predicted from the phase-space structure (see our Figure \ref{fig:blfd_phase_space}) and the power-spectra (see our Figures \ref{fig:power_z0} and \ref{fig:power_z_range}). \citet{Schekochihin2004} pointed out quite early that the power-spectra alone might not suffice to clearly identify dynamo action. Thus they suggest a different (stronger scaling) based on the curvature of the magnetic field lines given via: 
\begin{align}
    \mathbf{K} = \frac{(\mathbf{B} \cdot \nabla) \mathbf{B}}{|\mathbf{B}^2|},
\end{align}
which can be re-written as:
\begin{align}
    \mathbf{K} = \frac{1}{\mathbf{B^2}} \left[\frac{1}{2} \nabla (\mathbf{B} \cdot \mathbf{B}) - \mathbf{B} \times (\nabla \times \mathbf{B})\right].
\end{align}
We calculate this quantity as an additional output field in the code on the fly. We show the relation of the magnetic field strength as a function of the curvature of the field lines in Figure \ref{fig:curvature_plot} and note that while we are slightly too steep in the higher resolution models $10$X and $25$X we recover the declining trend of the field strength with the curvature following roughly $KB^{1/2} = constant$. Generally, this is a good sign as this indicates that the increasing field strength is counter-acting the bending of field lines by magnetic tension. Thus, the bending of field lines is suppressed by magnetic tension and the dynamo saturates in the regime of a few $\mu$G, as expected from the theory of the small scale-turbulent dynamo. The fact that our results are too steep could be related to our slightly too high magnetic field strengths in the cluster centre. Therefore, this could in our case be related to the some limitations of the model which we will discuss in detail in section \ref{sec:limitations}. Nevertheless, we raise two additional points about the curvature. First, this relation is mainly inferred from high-resolution plasma physics simulations without the presence of self-gravity. Thus it is a priori not clear why a galaxy cluster would exactly follow this relation as the gravitational collapse of structure will add an additional imprint on the curvature relation. Moreover, we note that this is not so different from \citet{Vazza2018} who are the only other group who ever checked this relation, where they also find a slight deviation from the results of \citet{Schekochihin2004}. We strongly suspect that the gravitational collapse of structure is responsible for the change in the magnetic curvature relation. However, we cannot proof this statement as this requires a detailed study of high resolution plasma physics simulations like the ones of \citet{Schekochihin2004} that include the effect of self-gravity.  

\subsection{Probability Distribution of the magnetic field}
\begin{figure}
    \centering
    \includegraphics[scale=0.33]{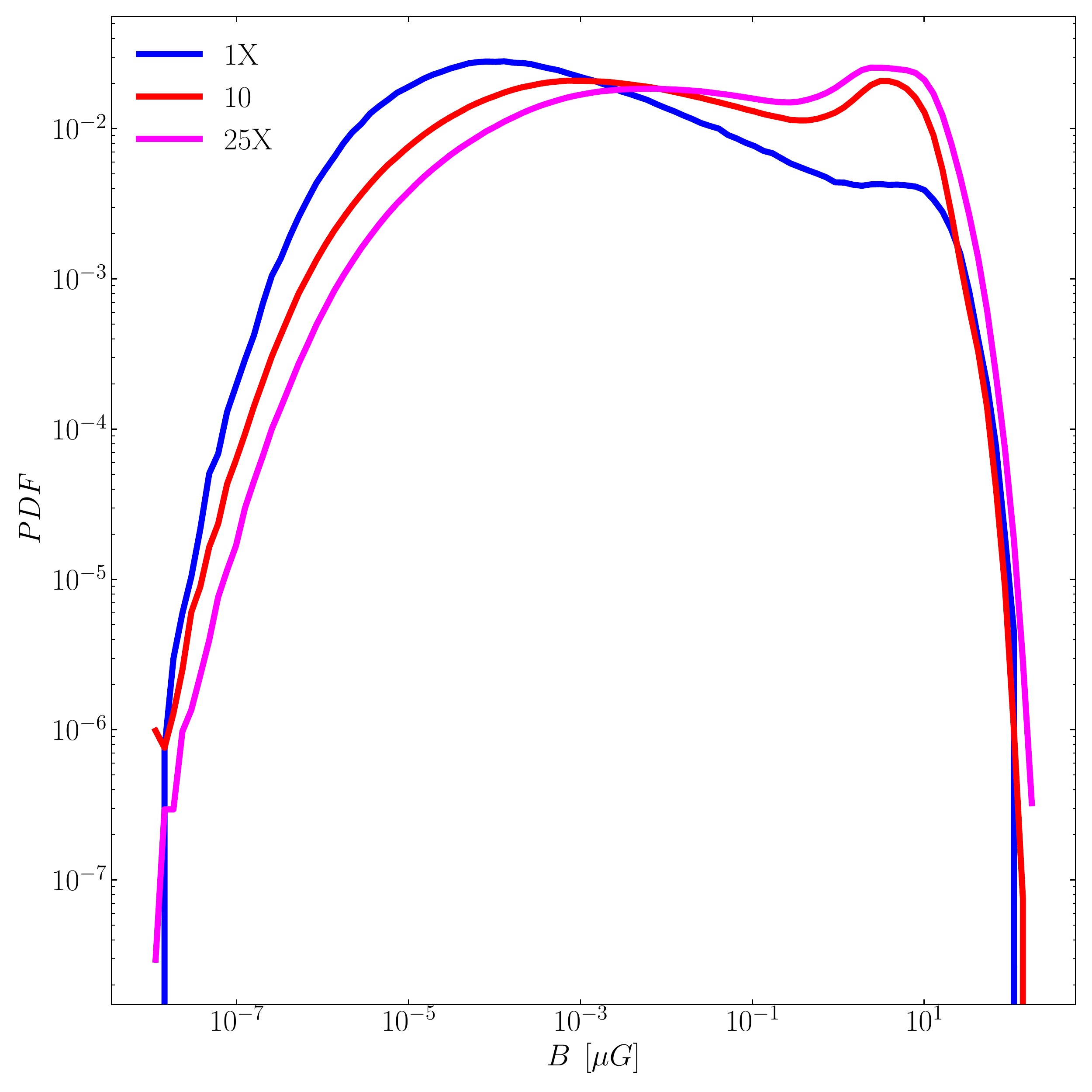}
    \caption{We show the volume weighted probability distribution function (PDF) of the density for all three models $1X$ (blue), $10X$ (red) and $25X$ (magenta), within the virial radius of our simulated galaxy clusters. For the lowest resolution model $1X$ there is a small plateau around the $\mu$G regime. In the higher resolution runs at $10X$ and $25X$ this evolves towards an extended bump showing a strong magnetisation of around $20$ per cent of the governed volume of the system. There is a sharp drop of the PDF for stronger magnetic field strengths above $10$ $\mu$G. This indicates that only very few gas cells govern the regime of extreme magnetic field strengths. Its interesting to point out that at low resolution the PDF peaks at relatively lower field strengths of around $10^{-9}$ indicating very inefficient dynamo amplification at lowest resolution level as the turbulence is not strong enough to significantly amplify the field into the $\mu$G-regime.}
    \label{fig:pdf}
\end{figure}

\begin{figure}
    \centering
    \includegraphics[scale=0.45]{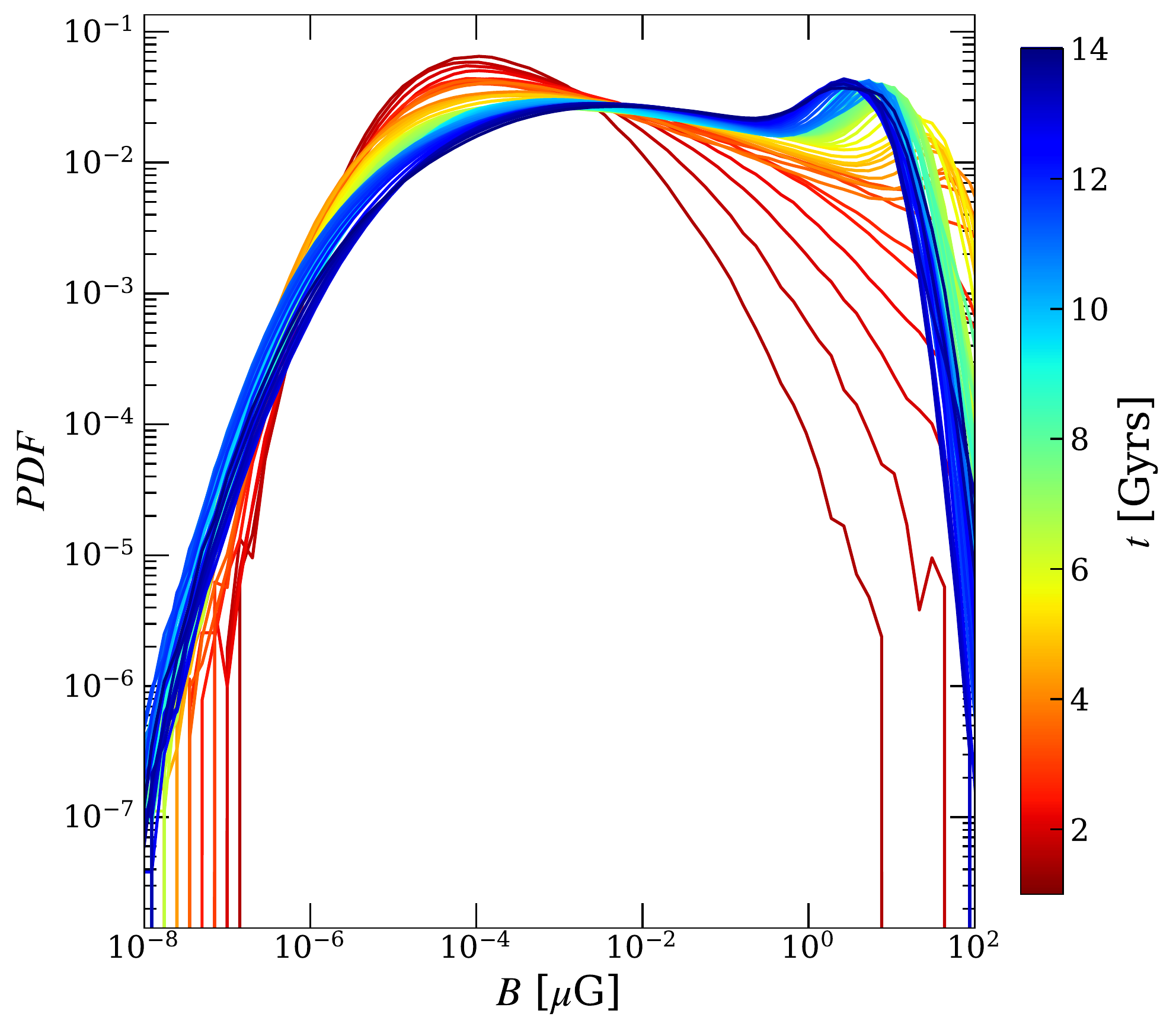}
    \caption{We show a time evolution of the volume weighted PDF of the magnetic field starting after $1$ Gyr of evolution of the cluster for our highest resolution run. Initially, we find a lot of the gas at very low field strengths around 10$^{-10}$ as this is corresponding to our initial value of our physical field. The dynamo action that is induced due to the turbulence introduced by structure formation is then amplifying to magnetic field as time is progressing forward. This redistributes the volume filling phase of the field to the higher magnetic field regime that is generated by the dynamo with decreasing redshift. Furthermore, the PDF is showing higher field values at redshift $z \sim 2$ compared to $z \sim 0$.}
    \label{fig:pdf_time_evo}
\end{figure}

Another aspect that is interesting when it comes to magnetic fields in galaxy clusters is the probability distribution of the field strength within the cluster. In this context there are two very important questions to answer. First, how does the magnetic field distribution change with resolution of the simulation and second, how does it change as a function of time within the cluster-region. \\
In Figure \ref{fig:pdf} we show the magnetic field PDF at redshift $z=0$ for our cluster at the three targeted resolution levels $1$X (blue), $10$X (red) and $25$X (magenta). In the lowest resolution run at $1$X we can see that the magnetic field distribution peaks at a low value between 10$^{-5}$ and 10$^{-4}$ $\mu$G, even at redshift $=0$ and there is only very little fraction of the volume that is reaching magnetic field up to a few $\mu$G. This picture changes in the higher resolution runs, where we can indicate a small peak between $1$ $\mu$G and $10$ $\mu$G, that shows that roughly ten percent of the clusters total volume reach significant magnetic field strengths that are in agreement with observed magnetic fields in galaxy clusters.  \\
Furthermore, we take the time evolution of the PDF of the magnetic field into account by displaying a time sequence of the PDF for our simulation $25$X that is colour coded against cosmic time starting from an evolutionary state that marks $1$ Gyr. We show this in Figure \ref{fig:pdf_time_evo}. Initially, the magnetic field distribution peaks at a value below 10$^{-4}$ $\mu$G, which is slightly higher then our initial seed-field value for this run due to adiabatic compression of the gas during structure formation. At later times on can identify a clear shift in the PDF from low magnetic field values to high magnetic field values where the peak is shifted the furthest to the right after around 4 Gyrs of evolution, which is roughly consistent with the slight peak in the mean magnetic field strength that we could observe for our higher resolution runs in Figure \ref{fig:bfld_evo} at around or slightly before a redshift of $z=2$. The peak shifts further to the left again with decreasing redshift and peaks at a value of around $7 \mu$G by redshift $z=0$. Furthermore, we note that the by combination of Figure \ref{fig:pdf} and Figure \ref{fig:pdf_time_evo} the dynamo action can be identified by a transition of the volume weighted PDF that peaks around the initial seed field value, at high resolution the field is redistributed to higher field strengths in the $\mu$G regime by the dynamo, while at low resolution the volume filling phase remains at the seed field value, even at redshift zero due to unresolved dynamo action. 

\subsection{Divergence cleaning constraint}
\label{sec:divb}

\begin{figure*}
    \centering
    \includegraphics[scale=0.35]{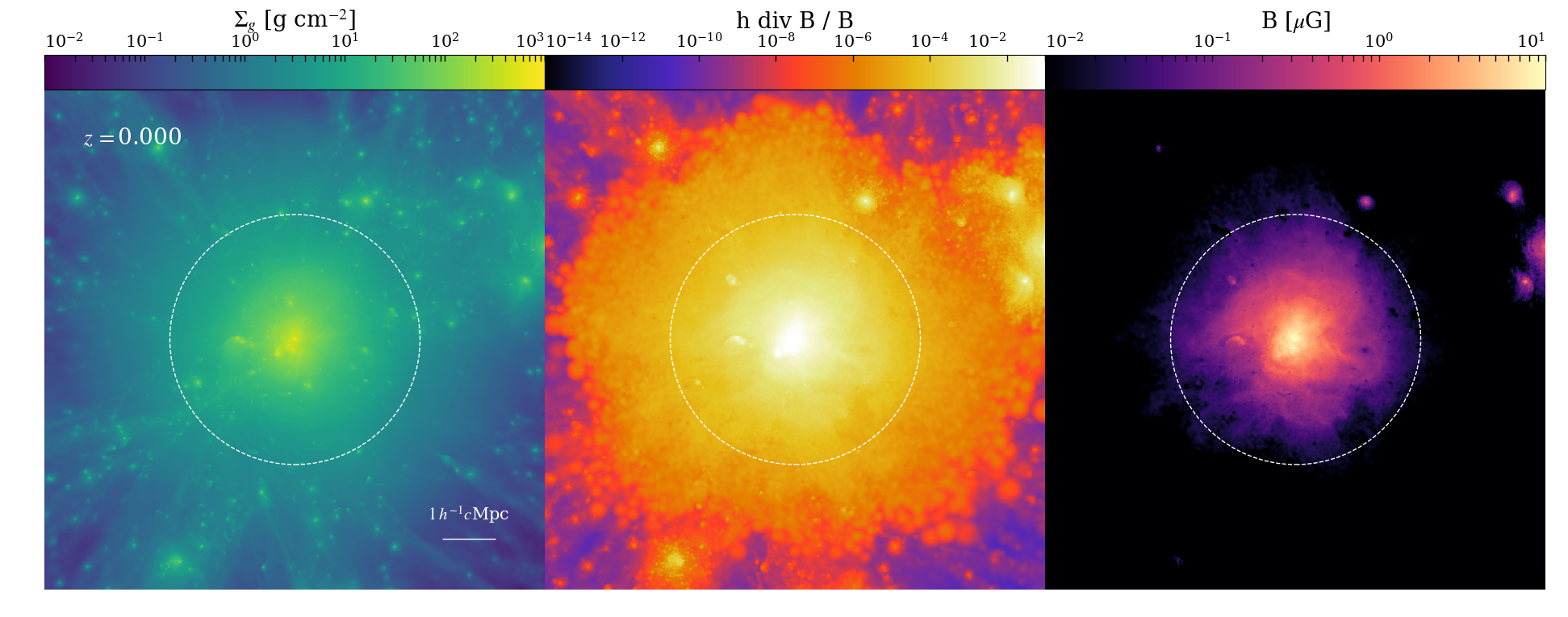}
    \caption{We visualise the gas surface density on the left, the divergence of the magnetic field in the centre and the magnetic field strength on the right for our highest resolution simulation $25$X. We note that our highest magnetic field strengths in the centre are associated with the highest divergence.}
    \label{fig:divB_cluster_color}
\end{figure*}

\begin{figure}
    \centering
    \includegraphics[scale=0.4]{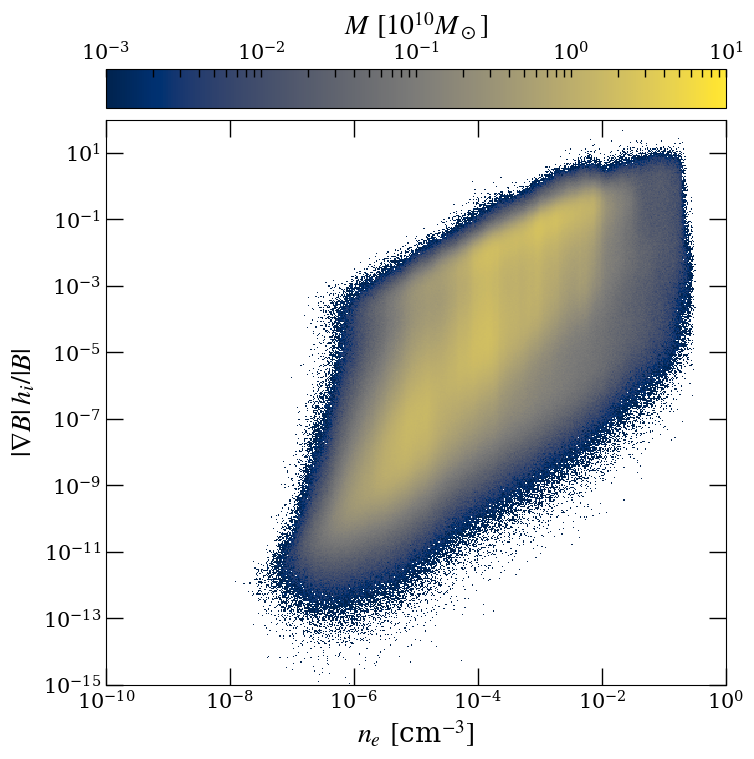}
    \caption{We show the density-divergence phase space colour coded against the mass in the bin each bin. Our simulations show excellent divergence cleaning properties. We show the innermost $2$ Mpc of the most massive structure within our simulation. One can clearly see that the bulk of the material shows excellent divergence cleaning properties with a relative divergence between $10^{-11}$ and $10^{-5}$. However, we note that single particles reach very high relative divergence around $10^{-1}$, while there are outliers reaching up to $1$. Despite this the excellent relative divergence makes it very unlikely that the magnetic field is solely amplified by magnetic monopoles.}
    \label{fig:divB_color}
\end{figure}

\begin{figure}
    \centering
    \includegraphics[scale=0.32]{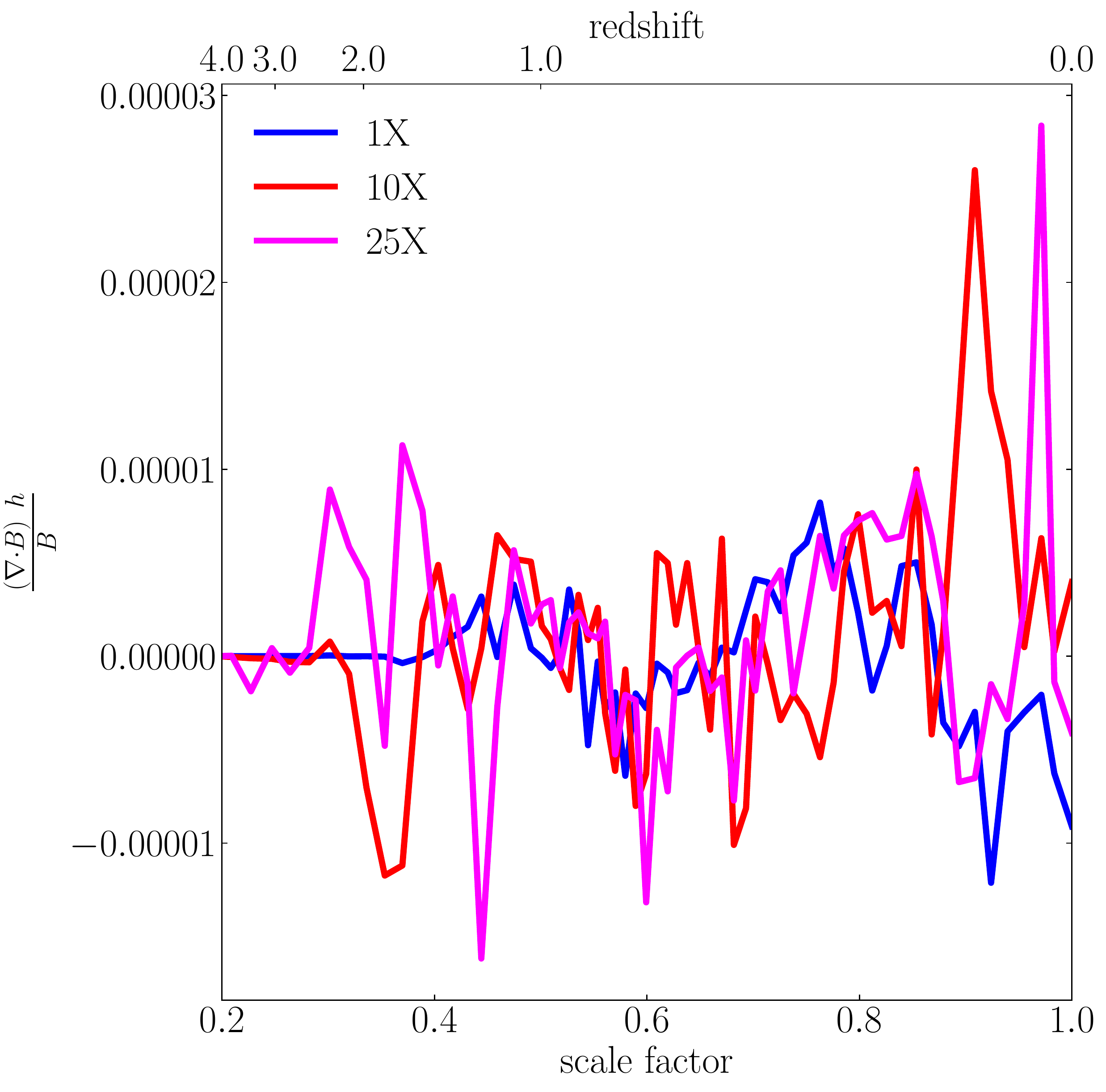}
    \caption{We show the time evolution of the mean value of the divergence for all three resolution levels of our galaxy cluster simulation with 1X (blue), 10X (red) and $25$X (magenta). We can clearly see that the divergence of the field remains very low until the cluster starts to grow a significant field around redshift $4$. The divergence fluctuates around the zero with peak values reaching $10^{-5}$ which is an excellent value compared to other galaxy cluster zoom simulations of the same kind \citep[e.g.][]{Vazza2018}.}
    \label{fig:divb_evo}
\end{figure}
Finally, we discuss the divergence constraint that is of importance for MHD simulations with both particle and grid based methods. While in grid based codes it is possible to enforce the divergence constraint by using the constrained transport (CT) scheme in which one is computing the cell centred magnetic field from the electric field on the edges of each cell, it is not yet clear if this can be done in a similar fashion in particle codes. Regarding a CT-scheme, one should keep two constraints in mind. First, strictly speaking with a CT-scheme one is enforcing that the numerical realisation of a physical field is divergence free, not the actual physical field (i.e. the field is divergence-free in the projected grid geometry). Second, and more importantly the divergence error of the magnetic field in a simulation is directly tracing the accuracy of the integration of the induction equation within the simulation. Thus, if present, the non-zero divergence gives a direct estimate on the integration error of the underlying integration scheme used for the induction equation. This information that is at least partially lost in a CT scheme. While the first point is rather picky and strictly speaking this is true for every numerical realisation of any physical system, the second one is rather important as it effectively measures how well the numerical scheme is handling the complexity of the MHD equations. As pointed out above, it is not clear if such a scheme can be constructed in fully Lagrangian code and thus particle codes (such as ours) rely on divergence cleaning schemes to control the error introduced by a non-zero divergence. In our simulations we use an $8$-wave  \citet{Powell1999} cleaning scheme.  \\
In Figure \ref{fig:divB_cluster_color} we show the column density (left), relative divergence (middle) and the magnetic field (right) for our simulation $25$X at redshift $z=0$. While we can see that the high divergence regions track the regions with high magnetic fields, the mean of the absolute value of the divergence stays reasonably low. Nevertheless, we want to point out that the non-zero divergence is the potential origin of the high central magnetic fields that we observe in our higher resolution runs at redshift $z=0$, if we compare to recent observations of galaxy cluster magnetic fields.\\ Furthermore, we investigate the divergence of the magnetic field in the most massive structure within our simulations within the virial radius R$_\mathrm{vir}$ as a function of the density. We show our results in Figure \ref{fig:divB_color} for redshift $z=0$. In both Figures discussed so far, we measure the divergence as the relative divergence:
\begin{align}
\left(\nabla \cdot \mathbf{B}\right)_\mathrm{rel} = \frac{\left(\nabla \cdot \mathbf{B} \right) h }{\mathbf{B}},
\end{align}
where $h$ is the smoothing length of kernel in our SPH simulation. We solve for this quantity on the fly by computing:
\begin{align}
\left(\nabla \cdot \mathbf{B}\right)_\mathrm{rel,i} = \left(\nabla \cdot \mathbf{B}_{i}\right) \frac{h_{i} + h_{j}}{B_{i} + B_{j}} \cdot \frac{B_{i}}{h_{i}}.
\label{eq:rdib}
\end{align}
in our simulation code, to obtain the most accurate description of our divergence error within the simulations. We directly write this data into our simulation snapshots. We can see from Figure \ref{fig:divB_color} that our relative divergence error remains very low for our $25$X resolution simulations, with the bulk of our particles showing a relative divergence error around $10^{-9}$. This is a very good result and shows that our code is capable of handling the divergence constrains in astrophysical MHD simulations to a sufficient amount. However, we note that there is an extended tail of the relative divergence error reaching up towards one for single particles at very high densities. These are also tracing the particles with the highest magnetic field in our simulations the show a field strength of around $100$ $\mu$G. This fact could be related to the fact that our higher resolution simulations predict slightly too high central magnetic fields within the cluster. Still, given the distribution of the relative divergence error in our simulations we can rule out magnetic monopoles as the primary amplification mechanism of the magnetic field in our simulations.\\
Moreover, we want to gauge the evolution of the divergence constraint as a function of time by showing the density-divergence phase space for three different redshifts in \ref{fig:divb_evo} ranging from redshift $2$ over redshift one to redshift $0$ on the right. 
Finally, we note that the results for the divergence of the magnetic field that we obtain in our galaxy cluster zoom-in simulations are excellent, even compared to results obtained with grid codes like \textsc{enzo}, \textsc{ramses} and \textsc{arepo}, which can for example be seen by comparing our divergence-density phase-space in Figure \ref{fig:divB_color} with the results obtained by  \citet[][]{Pakmor2020}. The origin of this is of our improved divergence constraint compared to previous SPH simulations can be understood as follows. First, all of our simulations are carried out in a non-radiative fashion, which avoids the cooling driven collapse of high density regions within the ICM into galaxies, which typically would host the regions with the largest divergence in the simulation domain. However, even compared to the non-radiative simulations of s\citet{Vazza2018} we obtain a result that shows around an order of magnitude lower divergence in the radial trend, which shows that our simple Powell-cleaning scheme \citep{Powell1999} is sufficient to capture the emergence of the magnetic field in the ICM, at least in non-radiative simulations. Moreover, we note that the divergence is already heavily suppressed compared to older SPH-simulations just by adopting a modern form of SPH which makes use of the higher-order kennel first discussed by \citet{Wendland1995}. Essentially, one can imagine that the use of a higher order kernel in SPH is similar to increasing the convergence order of the code, while decreasing the spatial resolution.\\ 
Second, we derive the relative divergence of the magnetic field in an SPH-like fashion in our simulation code following equation \ref{eq:rdib}, which is the correct way of obtaining the relative divergence in SPH simulations. If the divergence is derived from the particle data alone by weighting it with the smoothing length and normalising it to the absolute value of the magnetic field it can be heavily over- or under-estimated depending on the exact averaging that is applied which is usually very sensitive to the outliers in the distribution. Deriving the divergence following equation \ref{eq:rdib} is not prone to outliers as it is a kernel averaged quantity. Therefore, we propose that in all particle based methods the divergence should always be calculated by equation \ref{eq:rdib} to avoid confusion in future SPMHD simulations and provide a cleaner comparison to eulerian codes who often calculate this quantity by a finite-difference technique \citep[e.g.][]{Vazza2018}.

\section{Conclusions} \label{sec:conclusions}
\subsection{Summary}
We present SPMHD simulations of a massive galaxy cluster with a total mass of $M_\mathrm{200} \sim 10^{15}$ M$_{\odot}$ as a resolution study on three different resolution levels $1$x ($\sim 10^{8}$ M$_{\odot}$ per cell), $10$x ($\sim 10^{7}$ M$_{\odot}$ per cell) and $25$x ($\sim 4 \cdot 10^{6}$ M$_{\odot}$ per cell). We investigated the structure, morphology and evolution of the cluster, focused on the amplification of the magnetic field via the small-scale turbulent dynamo and discussed the limitations. The main conclusions of this work are the following:
\begin{enumerate}
    \item With increasing resolution the central magnetic field strength in the core of the cluster increases by a factor of $\sim 3$ from the base resolution run at $1$X towards the highest resolution run at $25$X. We note that this is higher than results obtained with Eulerian methods \citep[e.g.][]{Vazza2018} and observations of the Coma galaxy-cluster \citep[e.g.][]{Bonafede2011}, but is still not an unrealistic value for cool-core clusters.
    \item We find a steep exponential increase of the magnetic field as a function of cosmic time that flattens at a sub-equipartition value at around redshift $z=2$ for our lowest resolution simulation $1$X from which it increases at a slower rate as the cluster reaches redshift $z=0$ with a field that remains slightly below the $\mu$G-regime. In the higher resolution runs $10$X and $25$X we find that the magnetic fields peaks at around redshift $z=2$ fields saturates at a value of around $\sim 2$ $\mu$G in our $10$X and $25$X models while it stay slightly sub-equipartition in our $1$X model (when one compares the mean in the virial radius at redshift $z=0$).
    \item The field increase towards higher redshift is consistent with predictions for \textsc{ska} from \citet{Krause2009}.
    \item We find strong evidence that the magnetic field is amplified by the small-scale-turbulent dynamo in the ICM, driven by turbulence introduced by mergers, shocks and cosmic accretion. For the first time, we were thus able to unravel the non-linear regime of the dynamo driven by gravo-turbulence in agreement with the recent theoretical model, developed by \citet{Xu2020}. Furthermore, we show evidence for the dynamo by the magnetic power-spectra that take the form predicted by \citet{Kazantsev1968} and investigate the dependence between magnetic field strength and field line curvature and find good agreement with the results of \citet{Schekochihin2004} and \citet{Vazza2018}. 
    \item Finally, we analysed the behaviour of the divergence constraint in our simulations and find that while the divergence of the field is increasing with increasing resolution (which is the potential origin of our slightly too large central magnetic field strengths in the cluster centre) it is in good agreement with results presented with state of the art moving mesh codes like \textsc{arepo} \citep[e.g.][]{Pakmor2020} on galaxy scales and with state-of-the-art grid codes like \textsc{enzo} on cluster scales \citep[e.g.][]{Vazza2018}. 
\end{enumerate}
\subsection{Model limitations}
\label{sec:limitations}
Finally, we want to briefly discuss the consequences of the limitations of our modelling. First, we carry out the simulations without the effects of cooling, star formation an feedback. This might be an important restriction as one could imagine that the magnetic field is first amplified during the collapse of halos to proto-galaxies and later via galactic dynamos in the galaxies itself, that could redistribute their inter-galactic magnetic field to the CGM and subsequently the ICM via galactic outflows, driven by AGN and SNe. Specifically, the lack of AGN feedback could be related to our too high central magnetic field strengths in our higher resolution runs $10$X and $25$X. In that scenario, the high magnetic fields in the centre could efficiently support accretion towards a central SMBH, centred in brightest cluster galaxy (BCG) that would have formed under cooling in the centre of the cluster. The accretion of material and the subsequent outflow of material from the AGN, could efficiently transport the magnetic field outward and contribute to the magnetisation of the void regions surrounding the cluster. Throughout all the simulations we carried out, we build up a $\sim \mu$G magnetic field in the cluster centre.  The redistribution of the magnetic fields from the centre to the outer parts of the cluster via AGN feedback could also potentially help in decreasing the fast drop of the radial profile of the magnetic in the outer parts of the cluster.  \\
However, the aim of our study is to investigate magnetic field amplification in the ICM by the turbulent dynamo, driven by turbulence injected via shocks during structure formation processes. In order to understand how magnetic fields are growing in a cosmological context on large scales, we deliberately ignored these effects and ran the simulations in a non-radiative fashion.\\
Moreover, we didn't discuss the origin of the magnetic field in our simulations for example via the implementation of the Biermann-Battery or other battery processes driven by reionisation \citep[][]{Garaldi2020}. While the former could provide an interesting scenario for a self-consistent treatment of seed magnetic fields that are injected during the structure formation process by an offset between pressure and density gradient, the latter remain unimportant as long as we carry out the simulations in a non-radiative fashion. However, while we do not self-consistently implement a process like the Biermann-battery we tested different magnetic field strengths as initial seed for our $10$X runs with an order of magnitude difference and find very similar results. We discuss this in more detail in Appendix \ref{appendix:A}.
Additionally, we note that while we run all the simulations effectively in a non-ideal MHD limit we adopted a constant diffusion coefficient on all resolution scales. The change of the diffusion coefficient could have a potential effect on the magnetic field distribution, especially in the cluster centre where we find that the magnetic field strength is a factor of $2.5$ higher compared to observations of the Coma-galaxy cluster. We test the impact of a varying diffusion coefficient by increasing and decreasing by an order of magnitude in Appendix \ref{appendix:B}.
Furthermore, although we run MHD simulations we adopted isotropic conduction ignoring that magnetic fields can alter the thermal transport process in an anisotropic fashion. While we explicitly state this a caveat we tested the impact of anisotropic thermal conduction on basic cluster properties as well as basic dynamo properties and find only little differences when we include the effect for our $10$X and $25$X simulations. We discuss some of our findings in Appendix \ref{appendix:C}.\\
On top of this we also did not take other non-thermal effects like cosmic ray protons and cosmic ray electrons, which have a potentially important impact on the magnetic field in terms of amplification and structure \citep[][]{Buck2020, Butsky2020, Hopkins2020c, Hopkins2020b, Hopkins2020d, Hopkins2020a}.

\section*{Data Availability}
The data will be made available based on reasonable request to the corresponding author.
\acknowledgments
%
\vspace{5mm}
UPS is supported by the Simons Foundation through a Flatiron Research Fellowship (FRF) at the Center for Computational Astrophysics. The Flatiron Institute is supported by the Simons Foundation. UPS acknowledges computing time provided by the resources at the Flatiron Institute on the cluster rusty. UPS acknowledges the computing time provided by the Leibniz Rechenzentrum (LRZ) of the Bayrische Akademie der Wissenschaften on the machine SuperMUC-NG (pn72bu). UPS is acknowledging computing time provided by c2pap (pr27mi). KD, LMB and UPS acknowledge the computing time provided by the Leibniz Rechenzentrum (LRZ) of the Bayrische Akademie der Wissenschaften on the machine SuperMUC-NG (pr86re). This research was supported by the Excellence Cluster ORIGINS which is funded by the Deutsche Forschungsgemeinschaft (DFG, German Research Foundation) under Germany´s Excellence Strategy – EXC-2094 – 390783311. KD acknowledges funding for the COMPLEX project from the European Research Council (ERC) under the European Union’s Horizon 2020 research and innovation program grant agreement ERC-2019-AdG 860744. We thank the super computing resources at the LRZ in Garching for using an energy mix that is to $100$ per cent comprised out of renewable energy resources (e.g. \url{https://www.top500.org/news/germanys-most-powerful-supercomputer-comes-online/}, \\\url{https://www.lrz.de/wir/green-it_en/}).


\software{We use the cosmological simulation code \textsc{gadget3} \citep[][]{Springel2005, Dolag2009, Beck2016} to run the simulations and use the language \textsc{julia} \citep[][]{Bezanson2014} to perform the analysis\footnote{\url{https://docs.julialang.org}} based on the packages that can be found here: \url{https://github.com/LudwigBoess}}



\clearpage
\appendix
\section{Dependence on the seed field\label{appendix:A}}

\begin{figure}
    \centering
    \includegraphics[scale=0.32]{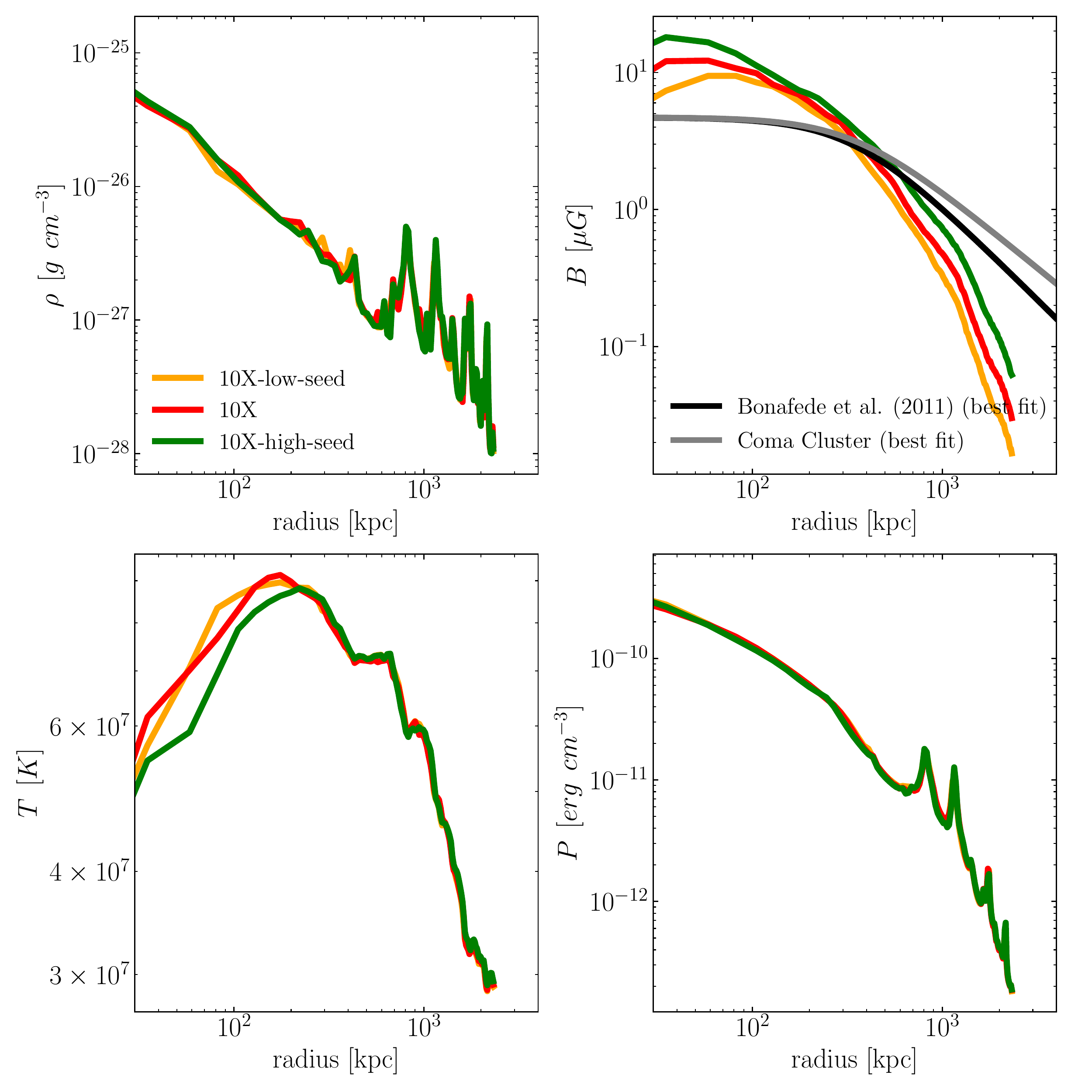}
    \caption{Same as Figure \ref{fig:radial_z0} but for the runs $10X$ (red), 10X-low-seed (orange) and 10X-high-seed (green) to demonstrate the effect of a variation in the initial seed field on the evolution of the cluster.}
    \label{fig:radial_z0_app_a}
\end{figure}

\begin{figure}
    \centering
    \includegraphics[scale=0.32]{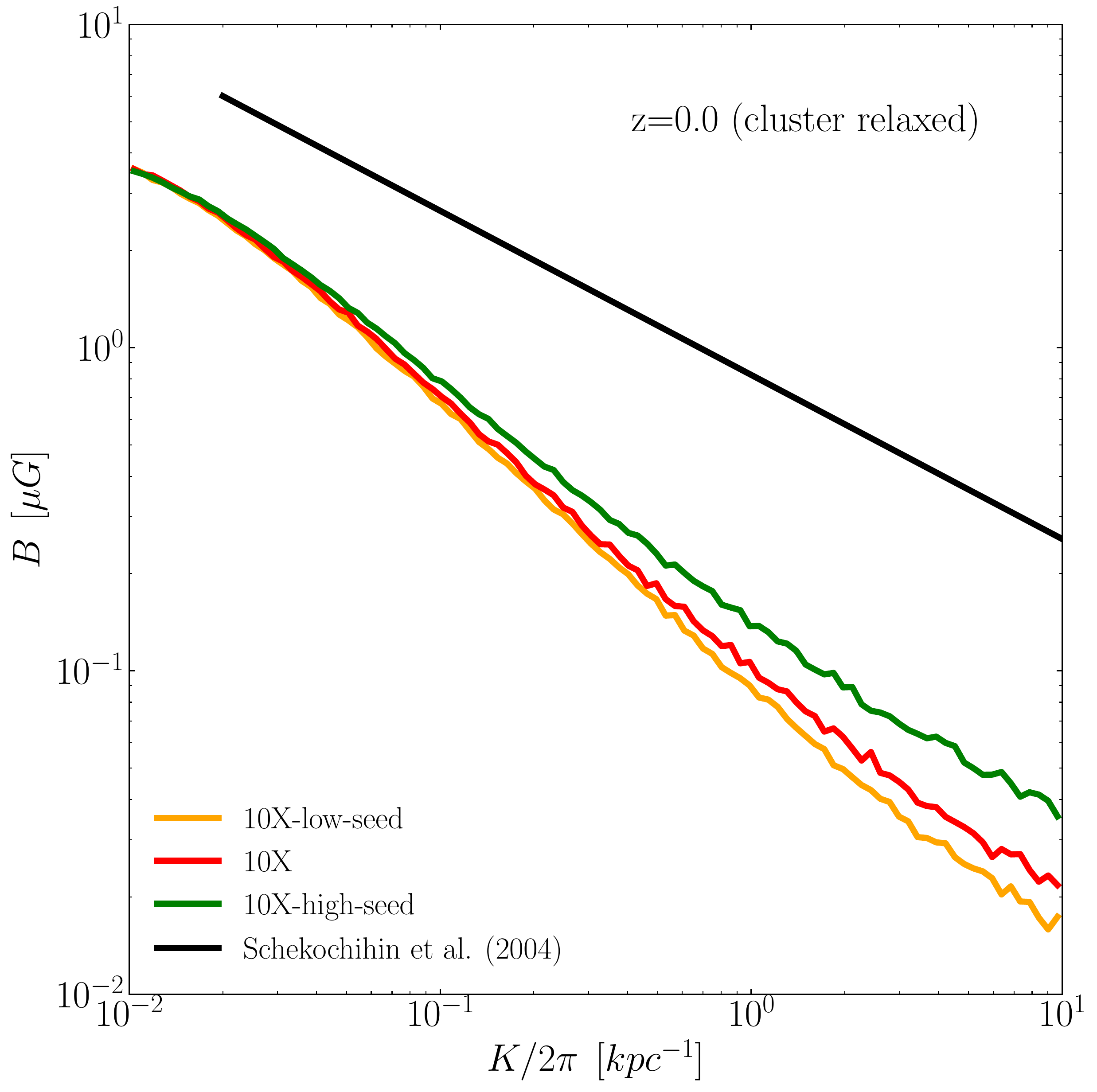}
    \caption{Same as Figure \ref{fig:curvature_plot} but for the runs $10X$ (red), 10X-low-seed (orange) and 10X-high-seed (green) to demonstrate the effect of a variation in the initial seed field on the evolution of the cluster.}
    \label{fig:curvature_plot_app_a}
\end{figure}

\begin{figure}
    \centering
    \includegraphics[scale=0.32]{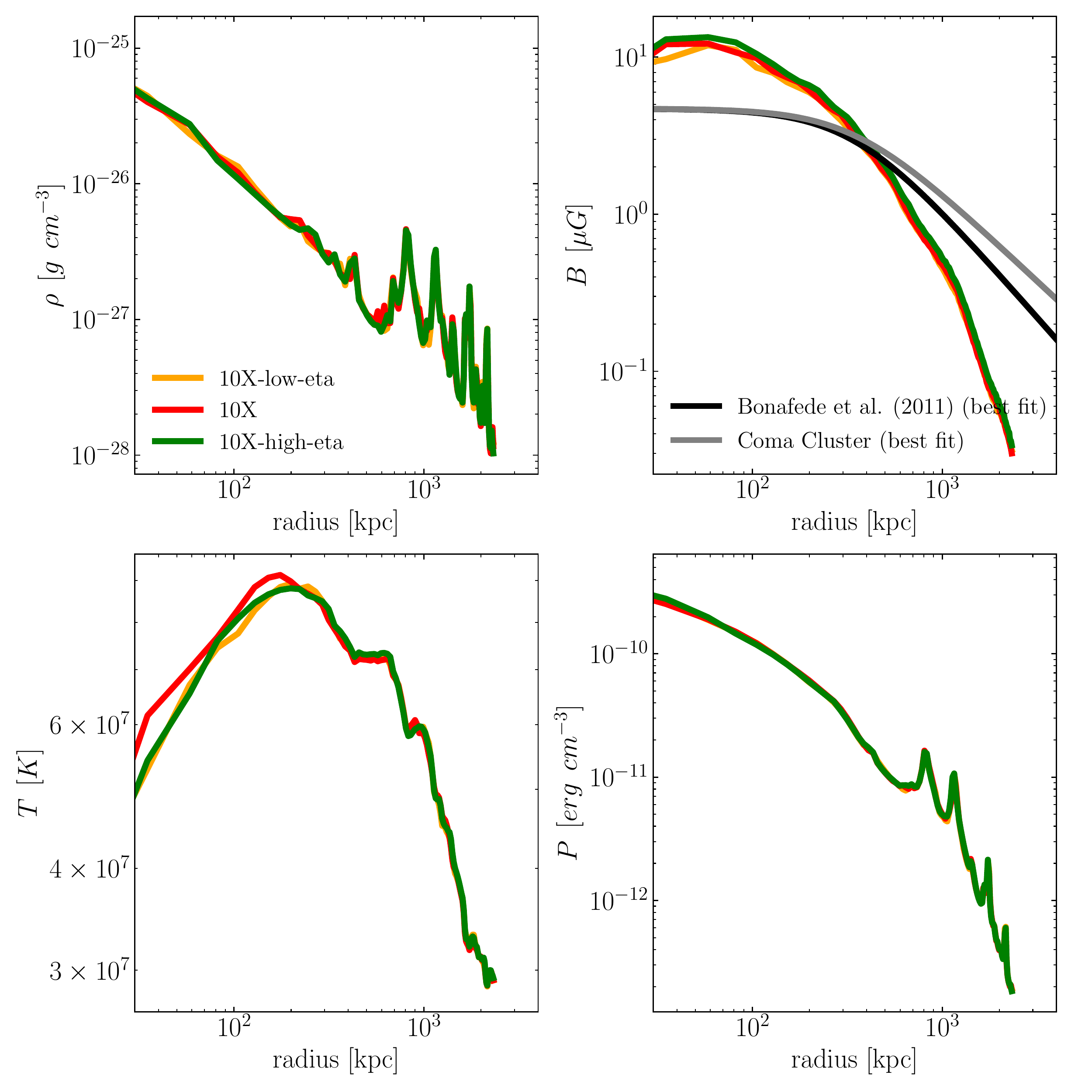}
    \caption{Same as Figure \ref{fig:radial_z0} but for the runs $10X$ (red), 10X-low-eta (orange) and 10X-high-eta (green) to demonstrate the effect of a variation in the magnetic resistivity of the plasma.}
    \label{fig:radial_z0_app_b}
\end{figure}

\begin{figure}
    \centering
    \includegraphics[scale=0.32]{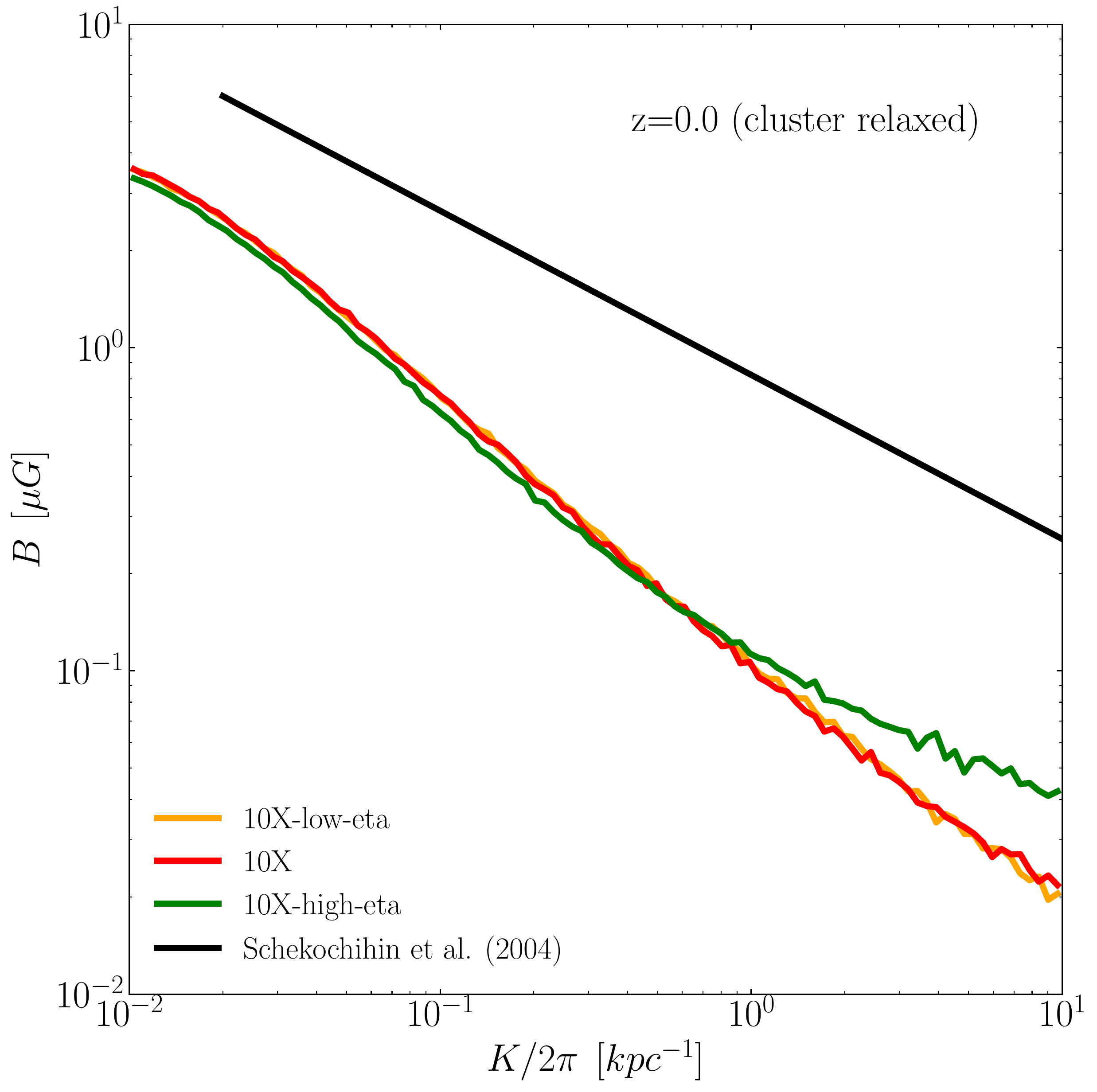}
    \caption{Same as Figure \ref{fig:curvature_plot} but for the runs $10X$ (red), 10X-low-eta (orange) and 10X-high-eta (green) to demonstrate the effect of a variation in the magnetic resistivity of the plasma.}
    \label{fig:curvature_plot_app_b}
\end{figure}

\begin{figure}
    \centering
    \includegraphics[scale=0.32]{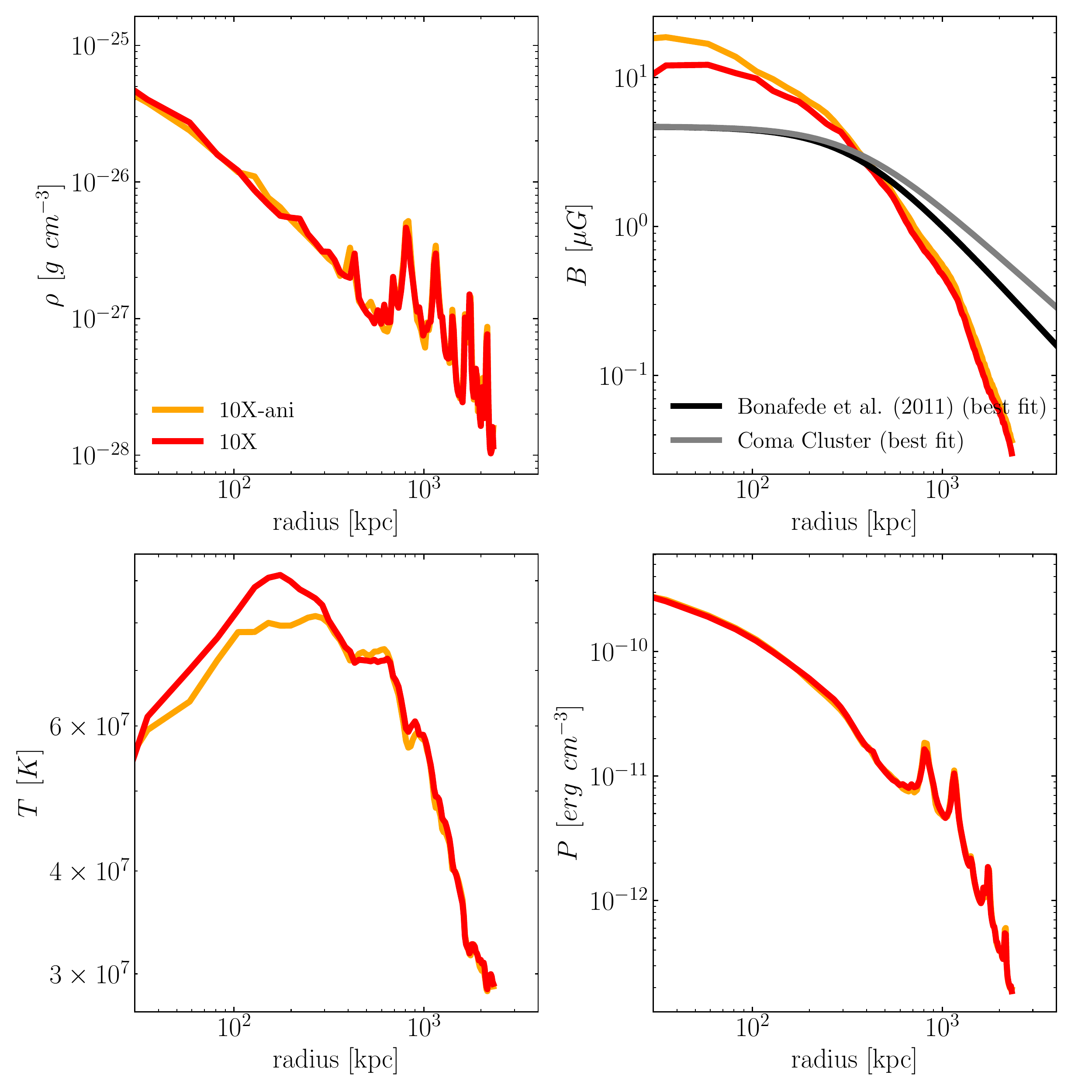}
    \caption{Same as Figure \ref{fig:radial_z0} but for the runs $10X$ (red) and 10X-ani (orange) to demonstrate the effect of anisotropic thermal conduction in the plasma.}
    \label{fig:radial_z0_app_c}
\end{figure}

\begin{figure}
    \centering
    \includegraphics[scale=0.32]{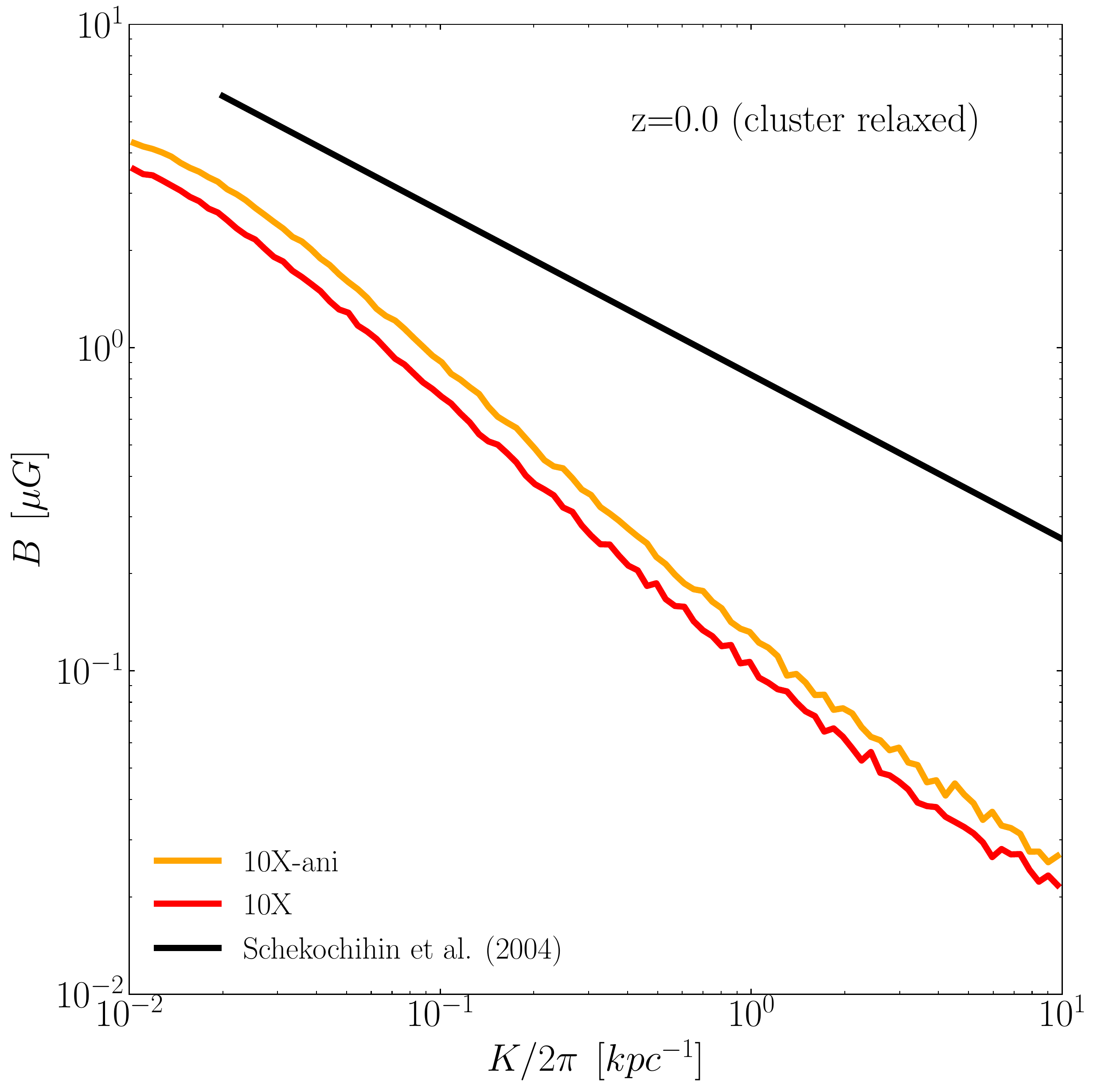}
    \caption{Same as Figure \ref{fig:curvature_plot} but for the runs $10X$ (red) and 10X-ani (orange) to demonstrate the effect of anisotropic thermal conduction in the plasma.}
    \label{fig:curvature_plot_app_c}
\end{figure}

The choice of the adopted seed field at high redshift is somewhat random and in fact its not our preferred choice of initialising magnetic field as we would much rather seed the field by supernovae. Generally, we think that the supernova-seeding scenario is physically better motivated than the choice of a pseudo-random seed field. However, as the runs are adiabatic, specifically to obtain a closer comparison to the work of \citet[][]{Vazza2018} we do not have cooling or star formation and cannot use our supernova-seeding approach that has been employed in earlier work \citep[e.g.][]{Beck2012, Steinwandel2019, Steinwandel2020, Steinwandel2020b}. Nevertheless, we can investigate and point out the key changes that we observe for a ten times higher and a ten times lower initial magnetic seed field. First, we note that the choice of the seed field does only marginally affect the magnetic field growth in most quantities. Thus we are mostly interested in the radial trend of the magnetic field which we find to be too high compared to the observations of Coma (see section \ref{sec:radial}). This is a known issue in Lagrangian methods \citep[e.g.][for a review]{Donnert2018} and it has been often suspected that this is related to the non-zero divergence behaviour in Lagrangian methods. Second, we are also interested in the magnetic field curvature relation as this is representing the direct imprint of the dynamo on the structure of the magnetic field lines, while power-spectra often take a very generic form and it remains unclear what drives their power law behaviour as we discussed in section \ref{sec:dynamo}. Therefore, these are the two quantities we want to focus on here. We show the radial profiles for density, magnetic field, temperature and thermal pressure in Figure \ref{fig:radial_z0_app_a}. There are two notable changes in the magnetic field structure that can be observed as the seed field is increased. The magnetic field grows roughly by factor of two in the very centre of the cluster with increasing seed field, while the variations in the outskirts of the cluster are marginal but we still find a slightly larger field. We note that while we find a relatively strong field for our run $10$X-high-seed with the higher seed field but also note that we find the lowest temperature in the cluster centre for this run. The drop in the temperature in the cluster centre is coming from the adiabatic nature of this set of simulations, possibly due to the absence of an AGN that provides energy to heat the gas.

Therefore we note that to increase of the cluster temperature in the centre alone would yield a drop of the central field by roughly half a dex. \\
In Figure \ref{fig:curvature_plot_app_a} we show the curvature relation as a function of the of the different seed fields and note that we find only very minor changes compared to our defaults runs. Nevertheless, the run with the higher magnetic seed field strength $10$X-high-seed captures the slope of \citet[][]{Schekochihin2004} better at larger magnetic field line curvature.

\section{Dependence on the magnetic diffusivity\label{appendix:B}}

Furthermore, we tested the dependence of our simulation results by adopting different magnetic diffusivity constants $\eta_\mathrm{m}$ in our non-ideal MHD prescription by increasing and decreasing the constant by a factor of $10$. We show the radial profiles that we obtain for these runs in Figure \ref{fig:radial_z0_app_b}. The differences are very minor and in fact they are contained within the model scatter at the per cent regime. This means that if we run the exact same simulation on a different machine or a different number of nodes we obtain a similar change in the radial magnetic field distribution as inferred from the change of the diffusion coefficient. This could be interpreted in the following way. The dynamo arises from the interplay between magnetic field amplification and dissipation (on small scales) or diffusion (to larger scales) of the magnetic field. The marginal change that we see by changing the diffusion coefficient tells us that the diffusion and dissipation of the field is of numerical nature rather than physical, which is consistent with other work that does not include the non-ideal MHD term to begin with (almost every other galaxy cluster simulation). However, there is an imprint of the change of the diffusion coefficient which is most apparent by considering the curvature relation in Figure \ref{fig:curvature_plot_app_b}. Here we can clearly see that the magnetic field is redistributed from the regime of smaller curvatures to larger ones, which is essentially forcing the curvature relation to become less steep at higher field line curvature.

\section{Dependence on the conduction model\label{appendix:C}}

Finally, we carry out one additional test on our $10$X model which is the inclusion of anisotropic thermal conduction. Generally, we note that the run with anisotropic thermal conduction produces the largest field in the radial trend of all the test runs (see Figure \ref{fig:radial_z0_app_c}). However, this can somewhat be understood by taking the temperature profile into account. We can clearly see that the run $10$X-ani shows the lowest temperature in the cluster centre and shows a consistent increase in the magnetic field compared to the decrease in temperature. Moreover, we find a very weak change in the curvature relation by the inclusion of anisotropic thermal conduction, which we show in Figure \ref{fig:curvature_plot_app_c}. On a side note we want to point out that these simulations are computationally expensive because we carry them out with physical conduction. The conduction module takes roughly $20$ per cent of the computing time which is a considerable computational effort. \\
In conclusion we note that the origin of the large magnetic fields in the centre that are too high by a factor of $2.5$ remains an unresolved issue in particle codes. We carefully checked that the divergence can almost be ruled out at this point as we show almost two orders of magnitude better behaviour compared to similar simulations that report magnetic fields that are a factor of $2$ too low compared to observations in the Coma cluster. Therefore, we need a more dedicated numerical study on the origin of the larger fields in Lagrangian codes, that will include model variations that include a more physics, such as Braginskii-viscosity. Moreover, we need to test more in depth the diffusion of the field and weather it is introduced by the cleaning scheme or the non-ideal MHD prescription. This will be subject of future work. 

\bibliography{sample63}{}
\bibliographystyle{aasjournal}



\end{document}